\documentclass[namedreferences]{solarphysics}
\usepackage[hyperref,optionalrh,showbiblabels]{spr-sola-addons} 
\usepackage{graphicx}                    
\usepackage{color}                       
\usepackage{breakurl}                    
\begin{document}

\begin{article}

\begin{opening}

\title{Always a Farm Boy}

%
\author{Brian R.~\surname{Dennis}
       }

%
\runningauthor{Dennis}
\runningtitle{Memoir}

%
  \institute{$^{1}$ NASA Goddard Space Flight Center\\
                     email: {brian.r.dennis@nasa.gov}\\ email: {brdennis@comcast.net}\\
             }

\begin{abstract}

This is the memoir of Brian Dennis - his life story to date, and his involvement with high energy solar physics.

\end{abstract}

%

\end{opening}

%
 \section{Introduction}
    \label{S-Introduction}

When Ed Cliver called me to say that I had been selected by the ``Memoir Committee'' to write an article for Solar Physics, I was in denial. Truth to be told, I was not aware of this series of memoir articles although I had read the first one by Kees de Jager - a delightful account of his early life in war-torn Europe and his involvement in some of the major advances in solar physics throughout his long and illustrative career. But why me?  I knew that there are many other people way better qualified than I to do this job, true giants in the field who have made major contributions to solar physics over their careers.  Why were they not asked to write about the great advances they had made over the last 40 to 50 years?  Well, in fact, many of them have already written wonderful articles for this series, well worth reading to learn how they came to their current state of understanding in their particular areas of expertise and what they see for the future. Indeed, the previous authors of papers in this memoir series - Jean-Claude Pecker, E. N. (Gene) Parker, Einar Tandberg-Hansen, Eric Priest, Hugh Hudson \citep{2016SoPh..291.1273H}, and now Peter Sturrock \citep{2017SoPh..292..147S} - make up an intimidating who's who list in the field.

So who am I to live in such exalted company? It turns out that I have had the privilege of working with some of those giants - people like Ken Frost,\footnote{https://en.wikipedia.org/wiki/Kenneth\_John\_Frost} Reuven Ramaty,\footnote{http://hesperia.gsfc.nasa.gov/hessi/news/ramaty.html} and Bob Lin.\footnote{http://news.berkeley.edu/2012/11/21/robert-lin-uc-berkeley-pioneer-in-experimental-space-physics-dies-at-70/}  Sadly they died way too young and are no longer with us to write the history of their contributions.  So, against my better judgment, I decided to accept the challenge to write up the history of high-energy solar physics as I have witnessed it, and to dedicate it to all the people who have made it such a rewarding experience for me, with those three gentlemen at the top of the list.  You, the reader, must decide if the Memoir Committee made the right decision in selecting me.  I will do my best to make this a worthwhile read as a relatively light-hearted romp through the history of high-energy solar physics as I saw it.


\begin{figure}
	\includegraphics[width = \linewidth]{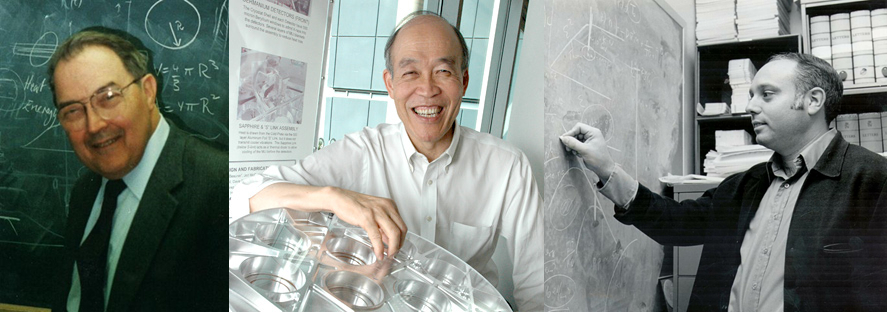}
    \caption{My three giants, Ken Frost, Bob Lin, and Reuven Ramaty.}
    \label{Fig-FrostLinRamaty}
\end{figure}

You can get a preview of what I am including in this paper by watching the video of the ``Maniac Talk'' that I gave at Goddard in 2014.\footnote{http://maniactalk.gestar.usra.edu/2014/10/from-picking-potatoes-to-measuring.html} This was one of a series of talks that are given once a month or so with the stated objective of promoting ``\textit{scientific interaction between young and experienced scientists in order to learn/improve/revise the knowledge of basics/fundamentals of science and scientific methods for research.}'' I called my talk ``From Picking Potatoes to Measuring the Biggest Bangs in the Solar System —- Always a Farm Boy!'' 

\section{Early Years in England}
    \label{S-England}

\begin{center}

\textbf{The Lincolnshire Poacher}\\\
When I was bound apprentice in famous Lincolnshire\\
Full well I served my master for nigh on seven years\\
Till I took up to poaching as you shall quickly hear\\
Oh, 'tis my delight on a shiny night in the season of the year\\
\end{center}


\begin{figure}
  \includegraphics*[width=1.0\textwidth, angle = 0, 
   trim = 0 0 0 0]
   		{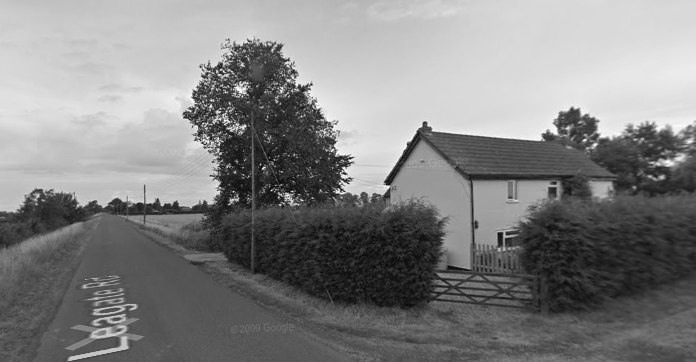}
    \caption{The house where I was born in Lincolnshire as it appears today. Note the one elm tree that gave its name to the house. The long straight road and the flat landscape are typical of this area called the Fens reclaimed from the sea in the 18th and 19th century.}   
	\label{Fig-TheElms}
\end{figure}

\begin{figure}
	\includegraphics[width=1.0\textwidth]{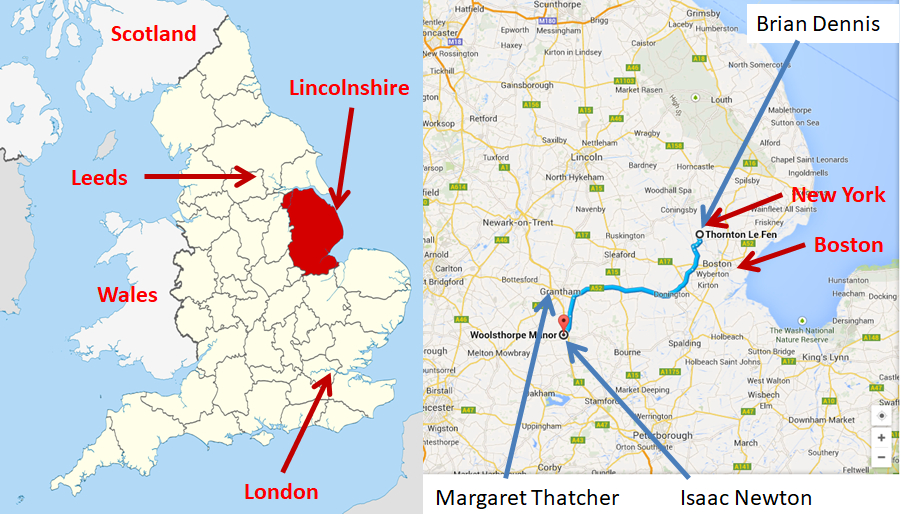}
    \caption{Map of England showing where I was born in Thornton-le-fen, near New York and Boston in Lincolnshire, England. Also shown are the birthplaces of the other famous people from Lincolnshire, Isaac Newton and Margaret Thatcher.}
    \label{Fig-Birthplace}
\end{figure}

I was born in 1940 in the house shown in Figure \ref{Fig-TheElms} and grew up on a 150-acre farm in ``famous'' Lincolnshire.  ``Where is that?'' people ask me in America.  ``Half way up on the right,'' I always answer, for truly it is on the east coast of England, about 120 miles north of London, as shown in Figure \ref{Fig-Birthplace}.  The farm was in a ``parish'' called, rather surprisingly, Thornton-le-fen, apparently a name left over from the Norman Conquest that followed the English defeat at the Battle of Hastings in 1066.  It is only a couple of miles from a village called New York and seven miles from the original Boston, so we had lots of reasons to feel connected to the US, even in those early days of the 1940's during Britain's finest hour as Winston Churchill put it.


A farm is perhaps the best place to be for a boy who is reasonably athletic and likes the outdoors.  I got to see nature in its rawest forms and had the freedom to explore, usually with little or no adult supervision.  Since there were not that many houses nearby, I mostly had just one friend who lived in the neighboring house about a quarter mile away.  His father did not own any land so he liked to come over to play football or cricket with me, go ``birds nesting'' or climb the trees in our orchard, or do the million other things that are possible on a farm.  He would come over to play just about every day so I was spoiled in this way and never learned how to ask other people to come play with me.  This was to be a problem when it came time to date girls, and even later in life, having to ask people to do things together.  I always like to wait until they ask me.

 Along with all these advantages of growing up on a farm, however, there was lots of often boring and hard work that was required at different times of the year. Fortunately, schooldays meant that I was spared many of the repetitive chores like milking the cow twice a day, feeding the cats and dog, feeding and cleaning out the bull, cows, and other cattle in the so-called ``crew yard,'' where they were housed over winter.  The latter task was particularly difficult since it required that you cut hay and straw from the stacks in the yard and carry great pitch-forks full into their stalls for food and bedding.  You also had to remove the ``soiled'' straw and put it on the ``muck heap'' to be spread later onto the fields as a fertilizer.  It was a great advance when we got a mechanical ``muck spreader'' driven by the tractor ``power take off'' so that we did not have to throw the composted material off a cart with a ``muck fork.''

 Perhaps the most enjoyable jobs on the farm was in the summer during harvest time.  The crops that we grew - oats, beans, barley, and wheat - were cut and bound into sheaves once they ripened.  The ``binder'' did this job with a 10-foot long horizontal steel blade running backwards and forwards to cut the long stems. A large rotating ``sail'' knocked the cut stems onto a conveyor belt that took them to be bound with ``binder twine'' into sheaves about a couple of feet in diameter.  The sheaves had to be propped vertically against one another on the ground by hand into ``stooks'' (pronounced ``stowks'' in Lincolnshire), each with a dozen or so sheaves.  This was to allow them to dry out and shed any water if it rained (which it usually did) before they could be carried to the yard on horse-drawn wagons.  The sheaves were either thrashed the same day to retrieve the grain or carefully arranged in huge 50-foot high stacks for later threshing.  My father was an expert builder of these stacks so that they did not (usually) fall over. Despite my fear of heights, I got to be good at pitch-forking him the sheaves one at a time with the twine knot down and the cut end out to make his stacking job easier.  He would often thatch the finished stacks with straw so that they were weather proof and could keep the grain dry over the winter despite England's famous somewhat wet climate.

 I liked to vary which of these several jobs I was doing as it could get very boring if you were stuck with the same job all day for days at a time.  When I was too young to do most of the harder work though, my main job was to lead the horses, one per wagon, from the yard to the field and back again. I would ride on the horse as it pulled an empty wagon down to the field to be loaded with sheaves and then lead another horse as it pulled a fully loaded wagon back to the yard.  This was a relatively easy job for me as the horses did all the work.

We had three horses - two strong young male horses called Punch and Prince, and their mother called Violet.  Violet was easy to handle and knew how to use her not inconsiderable strength in the most efficient way but her two boys were pretty headstrong and sometimes did not want to work. They would often yank at the harness instead of applying steady pressure to get the great 6-foot diameter wooden wagon wheels rolling. I got to enjoy working with all three of them and I think they got to respect me so that we got the work done without too much hassle.  Perhaps I learned about Newton's Laws of Motion from these horses and also how to get strong-willed people with big egos to do what needs to be done.  This experience has served me well in later life.

 An exciting few days on the farm would be when the threshing machine arrived.  This was a huge device to separate the grain from the straw of the oats, barley, or wheat.  It was operated by contractors who would go from farm to farm both during the harvest season and again in the winter.  The machine was made up of three parts, the ''drum'' about as big as a double-decker bus, the ``jacker'' or elevator about 50 feet long to carry the straw onto a stack that could get to be over 50 feet high, and a powerful tractor to power the whole thing using a belt wrapped around a rapidly rotating pulley wheel. The ``drum'' contained a large metal cylinder rotating at high speed (about 500 rpm according to Wikipedia) to separate the grain from the straw. A man on top of the machine would cut the twine holding a sheaf together and feed the contents into the drum with a big ``wumph.'' The straw was directed onto the jacker that would carry it up to a couple of people responsible for building the straw stack.  The ``chaff'' was collected at another location and hand carried into the barn for later use as animal feed and bedding.  The corn was directed to a chute at the front of the machine where it would be fed into large hemp bags.  A man would be assigned to tie the full bags weighing two hundredweight or two centum weights (cwt) or 16 stone or 224 pounds in the Imperial system.  He would then carry them into the barn, sometimes up a ladder to a second floor.  I could never lift such a weight but my brother, six years older than me, claimed that he once killed a rat while carrying one of these bags.

 If we were ``leading and threshing'' as they called it during harvest time, then my job would be critical to keep the fully loaded wagons arriving at the threshing machine at regular intervals so that it was never idle.  We had two wagons and a moffrey,\footnote{https://aipetcher.wordpress.com/tag/wheelwright/} a two-wheeled cart with two wheels and a wooden platform added on the front to give it a similar load carrying capacity of a wagon but with a smaller turning radius.
 
\begin{figure}
	\begin{center}
    \includegraphics*[width=0.8\textwidth, angle = 0, 
   trim = 0 0 0 0]
   		{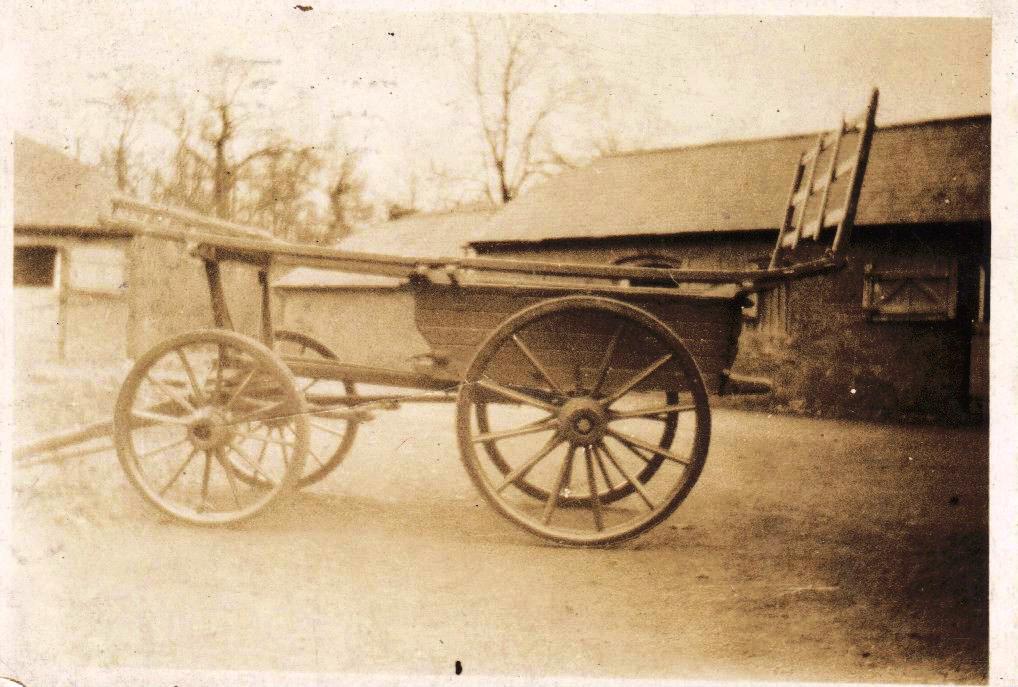}
    \caption{A specialist cart of East Anglian or Lincolnshire design called the `hermaphrodite,' and known locally as a moffrey.}
   \label{Fig-moffrey}
   \end{center}
\end{figure}

 Our three horses would all be in harness, yoked up one to each wagon or moffrey (Figure \ref{Fig-moffrey}).  While one wagon was being loaded in the field, and another was being unloaded into the threshing machine in the yard, I would be taking the third empty wagon down the field or bringing a full wagon back.  This carefully choreographed program would start at about 7 am, stop for a half hour at 11:30 for lunch, and continue until 4 pm for tea time.  Sometimes we would have another session from 5 'till 8 pm - a pretty long and tiring day.

 Some of the practices we indulged in on the farm will seem to be cruel and brutal when compared to today's attitudes towards animals but they seemed perfectly acceptable and necessary at the time. They left me with a very pragmatic attitude and a practical and totally unromantic relationship with nature.
 

Although I was born less than a year after Britain declared war on Germany in 1939, we were spared most of the horrors that other people suffered.  My father did not serve in the military, I believe, because there was an exemption from the draft for farmers but I am not sure of the details.  One of my uncles, also a farmer did not serve either.  Another uncle claimed conscientious objector status but he had to leave his home town of Boston, I later learned, because he was ostracized for refusing to serve.  None of this affected me very much and I grew up blissfully unaware of what was going on in the rest of the world.

We even escaped most of the effects of the strict food rationing, and always had plenty of food. We had cows on the farm so my mother made and sold butter and sour-milk cheese from the abundant milk.  We bought 100 day-old chicks every year so we had many chickens for eggs and a few cockerels for meat roaming about the yard all the time. We had a large garden and an orchard with over 20 trees so we had apples, pears, plums, raspberries, gooseberries, and many vegetables in season.

One of the main crops on the farm was potatoes.  Lincolnshire is famous for growing ``tates as big as yer 'ead'' translated as ``potatoes as big as your head.'' I well remember the days spent picking ``tates'' behind a horse-drawn ``spinner'' that plowed them up out of the ground and threw them onto the newly exposed soil. Back-breaking work. Many children took a week or more off school every autumn to pick the potatoes into ``mollies'' (large wicker baskets). Some of the potatoes were riddled (sorted by size) the same day and trucked in hemp bags each holding a ``hundredweight'' (112 pounds) to be sold in grocery shops or turned into crisps (potato chips in America) if the quality was high enough. The rest were stored over winter in ``tatey graves,'' 6-feet high Toblerone-shaped piles (i.e. with a triangular cross section) and insulated with straw and a foot-thick layer of soil over the top. I often helped in freezing weather later in the year to open the graves and riddle the potatoes to be sold off when the price was the highest. But we always saved enough potatoes for ourselves so that we had a continuous supply all year round.  It is then perhaps not so surprising that my mother served potatoes of one form or another - boiled, mashed, baked, fried, refried, etc. - at almost every meal.  They are still my favorite vegetable, especially ``new potatoes,'' the first potatoes ready in spring that my mother boiled with fresh mint from the garden and served with roast leg of lamb. We all thought she was the best cook in world and I could never understand why English cooking has got such a bad reputation world wide.

Another favorite time food-wise for me was in mid-winter when my father would kill a pig.  
I would usually take a day off from school to help my mother handle the meat from the pig.  One of my jobs was to grind the fresh pork and wind the handle of the screw device that we used to force the meat into the pig's intestines to make sausages.  My mother was expert at getting the ingredients - sage, rosemary, and thyme - in just the right proportions so that the sausages tasted just like the famous Lincolnshire sausages that you can still buy today.  But we had so much fresh meat and no way to refrigerate any of it since we had no electricity.  So we made up several plates with pork, sausages, hazelet, pork pies, and other delicacies and took them round to our neighbors to be eaten within a few days before they would spoil.  They returned the favor when they killed their pigs so we were eating fresh meat for most of the winter months.  My father salted down most of the hams in a big wooden box in the pantry to preserve them until later in the year when they could be soaked overnight and cooked.  The real delicacy was the main back ham which my mother would cut up into inch-thick slices with fresh parsley between and boil for several hours.  The resulting ``stuffed chine'' as it was called was everybody's favorite and can still be obtained throughout Lincolnshire.


Another annual experience that shaped my attitude towards life and animals in particular was the way we treated cats.  We always had several cats but they were not allowed in the house.  They were kept outdoors but with free access to the farm buildings.  They were well fed with table scraps and milk fresh from the cow twice a day. In return, they kept the mouse and mole population down. Twice a year, the mother cat would have a litter of several kittens but we only kept the ones born in the spring.  It was thought that the autumn cats would not well survive the winter and would not be as hardy.  
I delighted in making friends with the new spring kittens and stroking them, even when they were quite skittish as outdoor cats tend to be.  My wife was surprised when I showed considerable affection for the two kittens we acquired later in life. I enjoyed having them in the house, stroking them on my lap while watching television.

In addition to several cats, we had a sheepdog cross-breed named Gyp for as long as I can remember.  He was allowed in the house and was also fed on table scraps and milk.  He was not so good at herding the cows but he could run at up to 30 mph chasing cars along the road.  But his real value was as a rat catcher.  On winter nights, I would take him out to the farm buildings with a flashlight. As soon as I opened the barn door, he would dart inside and invariably catch a rat.

\section{An English Education}
    \label{S-Education}

\begin{figure}
 	\begin{center}
    \includegraphics*[width=0.8\textwidth, angle = 0, 
   trim = 0 0 0 0]
   		{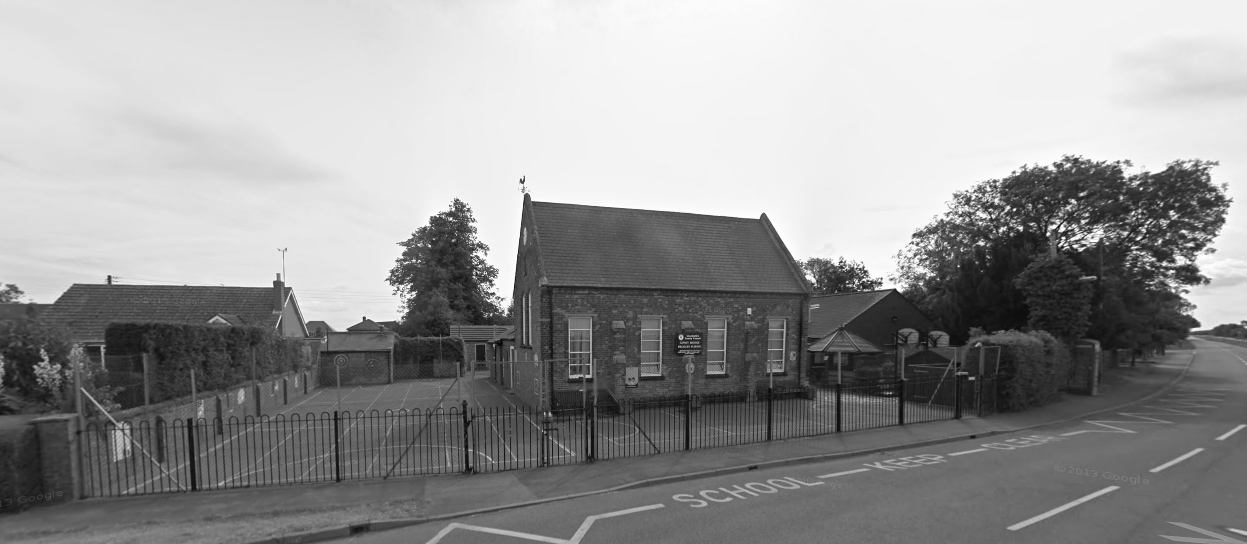}
    \caption{Gipsey Bridge Academy}
   \label{Fig-GipseyBridgeschool}
   \end{center}
\end{figure}

The parish where I was born and raised, Thornton-le-fen, had only about 200 inhabitants but it had its own school in the village of Gipsey Bridge about a mile from our farm.  In my day, it was officially called {\it Thornton-le-fen County Primary School} but it changed its name in 1956 to {\it Gipsey Bridge County Primary School} and is now called Gipsey Bridge Academy\footnote{http://www.gipseybridgeschool.co.uk/}.  This was an unimposing brick building with three classrooms surrounded by a concrete playground. The photograph in Figure \ref{Fig-GipseyBridgeschool} shows it as it is now but it has not changed much since my days there.

I started this school when I was 4 or 5 years old. I remember walking and later biking to the school along the dead-flat and nearly dead-straight road with just one slight $\sim10^{\circ}$ bend about half way along.  It always seemed to be head wind in both directions as there was little shelter except for the few trees along the way.

\begin{figure}
	\begin{center}
	\includegraphics[width=0.8\textwidth]{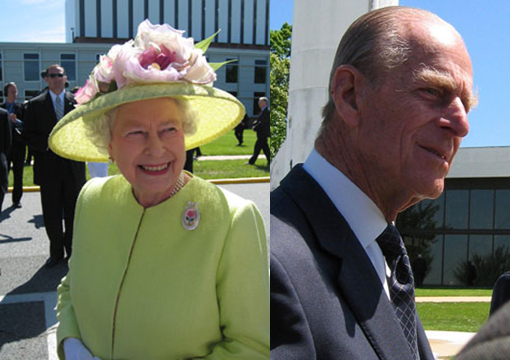}
    \caption{The Queen and Prince Phillip, the Duke of Edinburgh, on their visit to Goddard Space Flight Center in 2007.}
    \label{Fig-Queen_Phillip}
    \end{center}
\end{figure}

The main event of note during my time going to school at Gipsey Bridge was the wedding of Princess Elizabeth and the Duke of Edinburgh in 1948. There was a big celebration at the school that pretty well everybody in the neighborhood attended.  People loved it and always came out for all royal occasions.  The royal family was very popular then and can still draw huge crowds as evidenced by Lady Di's funeral and the popularity of Queen Elizabeth II and the whole royal family to this day. Little wonder that I love the new \emph{Netflix} series, \emph{``The Crown''} as I lived through all the events  covered in each episode.  I was a great fan of Prince Phillip, shown with the Queen in Figure \ref{Fig-Queen_Phillip}, and attribute much of my early interest in science to the example that he set. I have only seen him and the Queen once in person and that was in a limousine in Edinburgh as they arrived to open the Scottish parliament, coincidently just after a RHESSI workshop in Glasgow in about 2010.  

A critical step in my education was the so-called 11-plus test.  At that time in 1951, when I took the test, all children in the free schools in Britain had to take this test to decide if they would get an academic education at a grammar school or if they would go to a trade-oriented school, a so-called {\it secondary modern} school.  This test was not so controversial at the time and is still used in some parts of England today although it has been superseded by a more egalitarian system in most of Britain. I use the term {\it free school} since the term {\it public school} is used in Britain to denote the elite schools like Harrow and Eton that, apart from a few scholarships, have a hefty fee that only the aristocracy or the {\it nouveau riche} can afford. If you failed the 11-plus test, it was theoretically possible to switch from a secondary modern school to a grammar school after a few years but I know of very few people who were successful in doing that.  It is now easier to transfer, I understand, as shown by my grand-niece (my brother's granddaughter), Gemma Dennis, who managed to make the switch a few years after she failed her 11-plus in 2004.  Thus, at the time, it seemed that my only chance of getting into a free grammar school with the possibility of going on to a university was to pass the 11-plus exam.

\begin{figure}
 	\begin{center}
    \includegraphics*[width=0.8\textwidth, angle = 0, 
   trim = 0 0 0 0]
   		{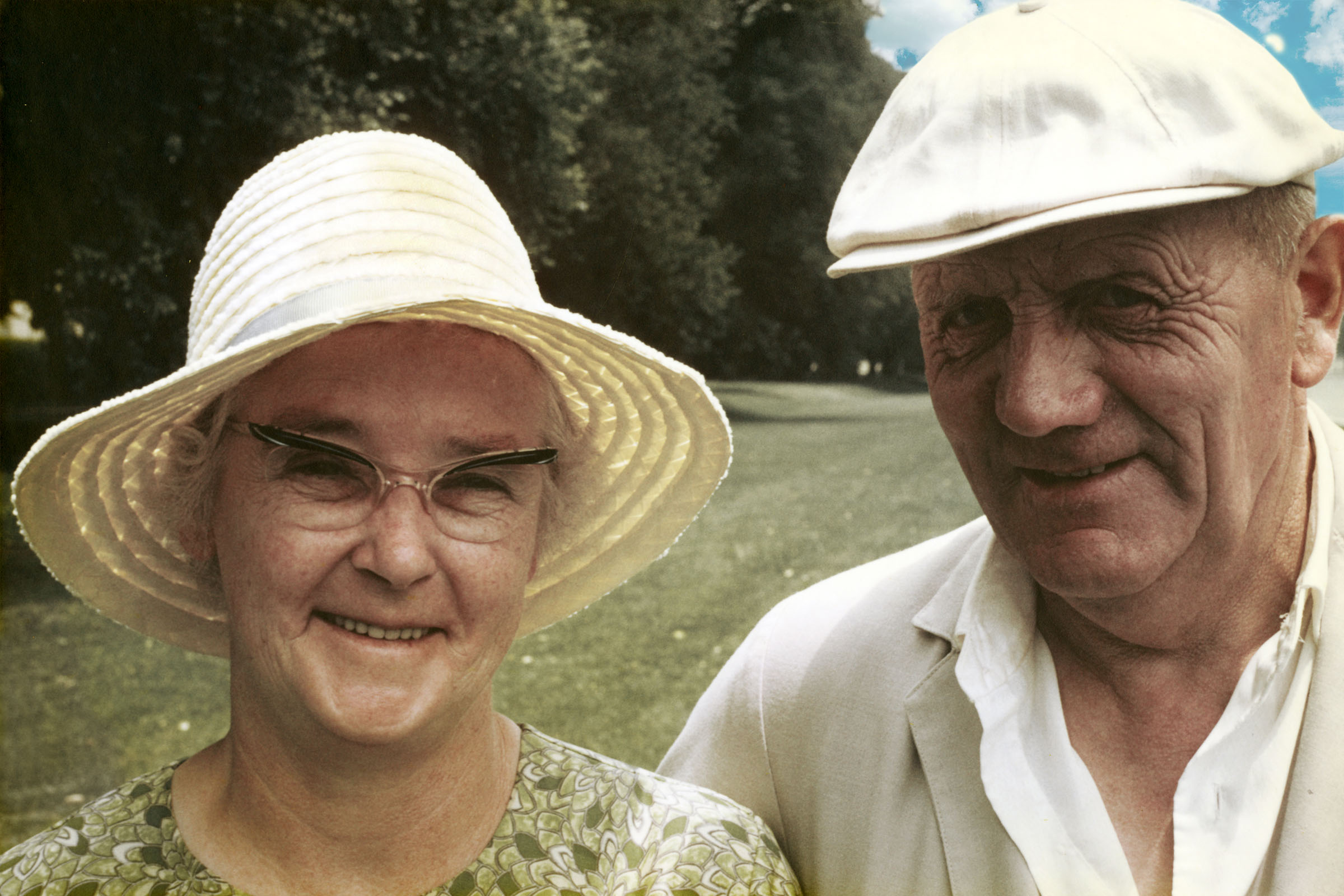}
    \caption{My mother and father on their one trip to America in 1969.}
   \label{Fig-MotherandFather}
   \end{center}
\end{figure}

I credit my mother, shown in Figure \ref{Fig-MotherandFather}, with getting me through the test. To my wife's amazement, my brother and sister and myself all remember our mother as the best friend we ever had.  She always took a keen interest in whatever we were doing, especially at school.  Apparently, she had wanted to marry a school teacher in the village but her father, the original owner of our farm, had forbidden it.  He did allow her to marry my father, a grocer's son from Boston who wanted to be a farmer but I suspect that made for a difficult relationship. My father was a {\it manager} at Gipsey Bridge School and, in fact, became {\it Chairman of the Managers} in 1958, so he was invested in our education but none of his children remember too many encouraging words from him.  He always seemed more interested in reading the {\it Daily Mail} newspaper during the week and his favorite, the ill-fated {\it News of the World} on Sundays, rather than talking to his children.

My brother, six years older than me, had failed the 11-plus exam and stayed on at Gipsey Bridge until he left school at age 16.  He was more the farmer type though and went on to attend Rhiseholme College\footnote{https://www.riseholme.ac.uk/} in Lincoln, where he first learned to take notes in classes on farming to go with the practicalities that he had learned at home. Following a stint in the army as a draftee going to Suez and Korea, fortunately just missing both conflicts, he ran the family farm very successfully after my father retired.  His elder son, Peter, is continuing the tradition.  Peter did pass his 11-plus exam but now says he would have been better off going to the secondary modern school since he was never interested in academics and much prefers the practical aspects of farming.

My sister, four years older than me, did pass the 11-plus exam in 1947 and went to \emph{Queen Elizabeth's Grammar School} \footnote{http://www.qegs.lincs.sch.uk/} in Horncastle, a small market town about 14 miles from our farm.  According to the booklet published to celebrate the 150-year anniversary of the Gipsey Bridge School founding in 1858, only eight children passed the 11-plus exam between 1945 and 1949. Nobody passed the year before I took it in 1951 and nobody passed it again until 1960, when two pupils passed including the headmaster's son.  So it was not a foregone conclusion that I would pass.  I do not remember any details about the test but I think it was an all-day written exam taken in the Gipsey Bridge school on a Saturday. Everything depended on that test.  In later years, candidates apparently took a preliminary test before being selected to take the main examination, with the final hurdle being an interview at Horncastle.

Well, I passed the 11-plus test and traveled the 14 miles by special bus to attend the same grammar school that my sister went to in Horncastle for seven years until I left for university in 1958. Of course, being the only pupil from Gipsey Bridge to pass the test that year, I knew nobody in my class at the grammar school.  It was very difficult to make friends since it seemed that everybody else already knew people from their primary schools.

We all had to wear the full school uniform including, for the boys, a maroon blazer, matching tie and peaked cap, grey shirt, charcoal trousers, and black shoes.  It was doubly difficult when on the very first day, I lost my cap.  Fortunately, my mother had sewn my name tag in every piece of clothing as required but it was thoroughly embarrassing when my cap was found and given to my sister, who was by now in the $4^{th}$~form.

I always struggled with school life.  Being the only pupil from Gipsey Bridge was bad enough but I soon found out that going to a {\it free} grammar school was also a badge of inferiority compared to pupils from the so-called {\it public} schools.  I well remember attending a group event organized by {\it Toc-H}\footnote{Toc-H is an international charity and membership movement that emerged from a soldiers' club in Poperinge, Belgium, during World War I. It works within the community to ease the burden of others and bring together disparate parts of society. http://www.toch-uk.org.uk/}. It involved about 20 boys from different schools around the country, both {\it public} and the free {\it grammar} schools.  We spent one memorable night in London going to many locations not generally seen by tourists, like Covent Garden at 5 am (while it was still the vegetable market made famous by the movie \emph{My Fair Lady}), the Smithfield meat market, and the publishing house of the Times newspaper (while it was still a national institution before it was bought out by Rupert Murdoch).  But the thing that sticks in my craw about that memorable trip was the comment made by one of the {\it public-school} boys when we were discussing what types of music we liked best.  I said, quite honestly, that I did not really like any music (and I still do not).  His comment was, ``Well, what do you expect from a {\it grammar-school boy}.''  For some reason, that put-down has remained with me to this day, and I am still trying to cope with it. I can only imagine how anybody who failed the 11-plus exam or indeed any of life's many tests, for that matter, copes with similar unthinking comments.

Well, I went on to do pretty well at Horncastle Grammar School, probably making my share of discouraging remarks to people less fortunate than myself.  At age 15, after the $5^{th}$ form, everybody took the so-called O-levels, the Ordinary Level of the General Certificate of Education (GCE). I passed in five subjects - English, history, mathematics (maths), geography, and chemistry.  I am embarrassed to say that I did not pass the French test on my first attempt, the only school test I ever failed.  This was undoubtedly because we had a very bad French teacher who tragically committed suicide, partly, I am sure, because of his lack of success in teaching us anything.  I also had dropped Latin the year before so I did not even take the O-level test in that subject.  This was to preclude any chance of going to Oxford or Cambridge University since, at that time, O-level Latin was a prerequisite for getting into either one of those prestigious places.

After two years in the $6^{th}$ form, the relatively few of us who had done well enough in the O-levels to stay on at school took the Advanced Level or A-level GCE.  I passed in the four subjects I took - Maths, Chemistry, Physics, and Advanced Maths.  We only got a physics teacher in the school when I entered the $6^{th}$ form so that is why I didn't take physics at the O-level.  We had to learn all of the physics needed to pass the A-level exam in just two years so it was a very concentrated class, fortunately with an excellent teacher. Perhaps this intense introduction to physics was the best way for me since it held my interest like no other subject before.  I was attracted to the common sense and practical aspect to it, at least until we dabbled in quantum physics, but Newton's Laws of Motion were a great awakening for me.  Amazingly,  our teacher did not point out that Isaac Newton was born only 30 miles from Horncastle at Woolsthorpe Manor, also in {\it famous Lincolnshire}.  I only recently discovered this fact, myself, and now tell anybody who will listen that there are three famous people from Lincolnshire - Isaac Newton, Margaret Thatcher born in Grantham, five miles from Woolsthorpe Manor, and myself.

What to do after grammar school?  My mother and I read up on the job opportunities that were available to a grammar school boy.  It seemed that the best bet was to go to a university and get a degree if you could get accepted. At that time only about 5\% of schoolboys went to a university and I would be the first from our family.  I applied to several of the so-called ``red-brick'' universities - London, Leeds, Sheffield, Manchester.  Oxford, Cambridge, and probably Durham were out because I did not have Latin. I was rejected by all except Leeds, who asked me to come for an interview.  They said that they thought I was too young, since my birthday was late in the school year and I was always one of the youngest in my class.  They recommended that I stay on at school for another year but that they would accept me if I wanted to come. With that resounding endorsement, I decided to go to Leeds and study physics.

\section{Leeds University (1958 - 1964)}

\begin{figure}
  \begin{center}
      \includegraphics[width=0.8\textwidth, 
      trim = 0 0 0 0]
      {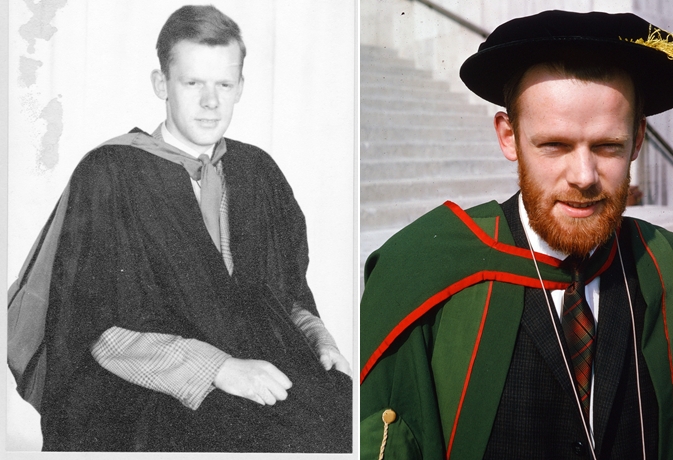}
      \caption{Me in my BSc gown on the left and in the somewhat more elaborate PhD outfit on the right.}
  \end{center}
  \label{Fig-BScPhD}
\end{figure}

My six years at Leeds University seem pretty uneventful now but at the time they were a real growth period in my life.  I spent three years taking classes in mathematics, physics, and various technical courses that I don't even remember anymore, leading to a second-class honors Bachelor of Science (BSc) degree in 1961.  The “second-class” label showed that I wasn't the smartest kid in the class but it was sufficient for me to be accepted to continue on as a graduate student specializing in cosmic ray physics.  At that time, it was very unusual for a student to switch to another university for a PhD so there was little question that I would stay at Leeds.  The choices were solid-state physics working in a lab in the basement, or trying to get a huge cloud chamber working that was the pet project of one of the professors, or working on a new extensive air shower (EAS) array that was being set up at Haverah Park on the Yorkshire Moors, just north of Leeds near Harrogate.  

I was very fortunate to be accepted to work on the Haverah Park air shower array as it suited my desire to get out of the city and to do experimental work.  It turned out to be similar to farm work but less back-breaking and much more interesting as we were doing new things that nobody else had done before. The leader of the project that had received funding a couple of years earlier was Professor J. G. Wilson, the head of the Physics Department at the time. My faculty supervisor was Dr. R. M. Tennent, but I rarely interacted with either of them.  Two graduate students, Ted Turver and Steve Lillicrap, had already spent a year working on it when I started, and a new and somewhat controversial faculty member, Harry Hunter, also joined the team to develop numerical methods for deducing the arrival directions and energies of primary cosmic rays from the EAS observations.  I owe them all a huge debt of gratitude for helping me get started in this field and becoming good friends in the process.  The project had its own Land Rover (before synchromesh!) and we drove the 17 miles out to Haverah Park almost every day.  We worked largely independently on first setting up the array and eventually taking data. Figure 9
shows me standing outside the central huts at Haverah Park, giving some idea of the bleak but beautiful location of this remote site.

\begin{figure}
	\begin{center}
      \includegraphics*[width=0.8\textwidth, angle = 0, 
      trim = 0 0 0 0]
          {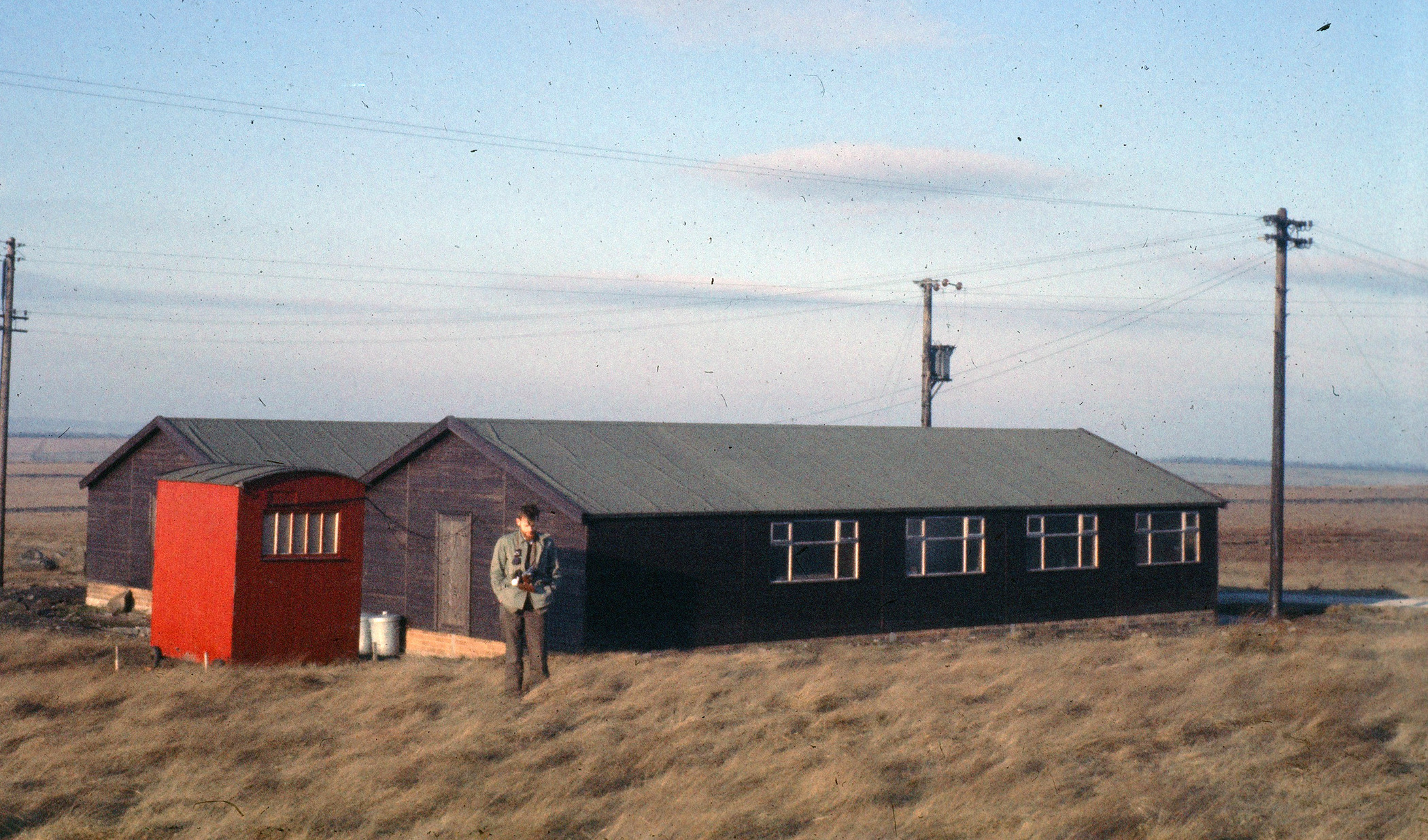}
      \caption{Here's me outside the central huts at Haverah Park on the Yorkshire Moors, about 15 miles north of Leeds.}
    \end{center}
   \label{Fig-HPoutside}
\end{figure}

The experiment at Haverah Park was designed to detect extensive air showers initiated by primary cosmic ray particles with energies of $\sim10^{17}$ eV and higher. Full details are given in my PhD thesis that, of course, is not available on line but I can lend you a copy if you really want to know.  A brief description is given in two papers by my nominal supervisor, Dr. Tennent \citep{1967PPS....92..622T, 1968CaJPS..46....1T}. Interestingly, there is no record in NASA ADS of any other papers with Tennent as lead author, except for a \textit{Nature} article on ``Cosmic-ray Showers'' \citep{1959Natur.183.1555T}. I didn't realize he had so few publications. (I noticed that in all publications at that time, including throughout my thesis, the abbreviation for electron volt was all lower case. I wonder when it changed to a capital V?)

The extensive air showers were detected at Haverah Park by an array of four Cerenkov detector stations covering an area of 0.7~km$^2$.  Each station consisted of 16 large iron tanks (6 ft x 6 ft x 4 ft deep) filled with Yorkshire's purest water shipped in by the local fire department.  The Cerenkov light produced by the shower particles -- primarily charged muons as they traveled through the water -- was detected by a single 5-inch diameter photomultiplier tube (PMT) mounted in the top cover of the tank. The signals from the photomultipliers were displayed on oscilloscopes and the traces photographed on 35 mm film. Figure 10
shows me inside the central hut with all the electronics and recording equipment. 

We had the job of setting up this whole system and keeping it going round the clock to record as many EAS as possible.  The full array became operational in December 1962, and by March 1964, when I started writing my thesis, we had recorded over 12 thousand events.  My thesis is primarily about analyzing this data set and determining, for most of the events, the arrival direction, the curvature of the shower front, the position of the shower core, and the total number of particles in the shower.  From this, the energies and arrival directions of about 6000 primary cosmic rays with a mean energy of $10^{17}$ eV were obtained and the results published in \cite{1963ICRC....4..258D}. No evidence was found for any anisotropy so no definitive advances were made in elucidating the origin of the cosmic rays, the main goal of the project.

\begin{figure}
	\begin{center}
      \includegraphics*[width=0.8\textwidth, angle = 0, 
      trim = 0 0 0 0]
          {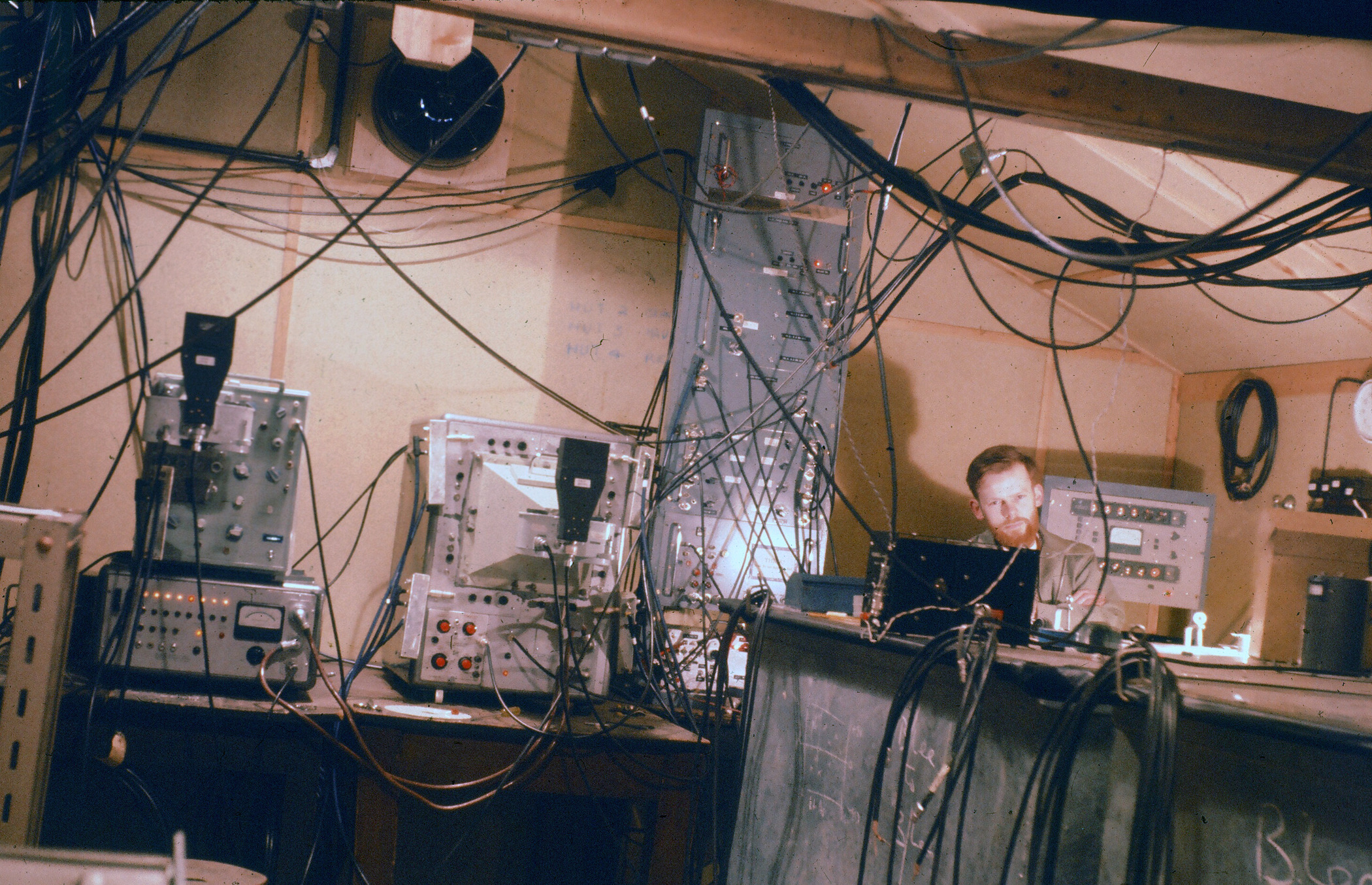}
      \caption{Here's me in the central hut at Haverah Park with all the electronics and recording equipment. One of the Cerenkov water tanks (6 ft by 4 ft by 4 ft deep) is in the right foreground.}
    \end{center}
   \label{Fig-HPinside}
\end{figure}





 Perhaps the later graduate students appreciated my work and benefited from the operational and data analysis techniques that we had developed. Sadly, the only citation to my paper with Professor Wilson was made 15 years later by \cite{1978JPhG....4..133E}. They state that their work incorporates much of the earlier material but ``in so far as we now apply consistent selection criteria to the whole energy range and group the showers in narrow energy intervals, the results presented here supersede the earlier work so that detailed comparisons with it are not appropriate.'' It is a little deflating to read a sentence like that but I guess all pioneers must feel this way when their work is ``superseded.'' The EAS array at Haverah Park was subsequently greatly expanded, extending the sensitive area from the 0.5~km$^2$ in my day to eventually 12~$km^2$ before it was turned off in 1987. Amazingly, it was reported that after over 20 years one of the detectors was opened and the water was found to be crystal clear. In fact it was good enough to drink.\footnote{
https://archive.is/20120722201836 \par
/http://www.ast.leeds.ac.uk/haverah/havpark.shtml\#selection-9.2485-9.2676}

It is gratifying to note that the closure of the Haverah Park EAS array was not the end of the story. A giant air shower array was built in Argentina with many aspects of the design of the array influenced by scientists from Leeds. In particular, one of the detector designs is based on the water Cerenkov tanks utilized so successfully at Haverah Park. This new facility covers 12 000~km$^2$ (the size of Lancashire!), and is known as
the Pierre Auger Observatory\footnote{https://www.auger.org/} in honor of the man who first discovered extensive air showers. 

I was sad to learn that Michael Hillas, who I had known at Leeds and who helped in the design of this observatory, died last year aged 85.\footnote{https://www.auger.org/images/PDFs/2017-11\_Obituary\_Michael-Hillas.pdf} It has been said that \emph{``if there had been a Nobel Prize for cosmic rays, Michael would have it,''} so I was surprised to learn from his obituary that he came to \emph{``Leeds in 1959 as a lecturer making it understood that he would not work on the embryonic air-shower experiment that was being developed under the leadership of Professor J G Wilson at Haverah Park.''} I wonder what that was all about.  

\begin{figure}
  \begin{center}
    \includegraphics*[width=0.7\textwidth, angle = 0, 
   trim = 0 0 0 0]
   		{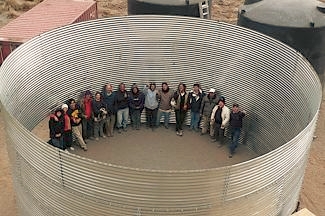}
    \caption{HWAC water Cerenkov tank, 4 meters high and 7.3 meters in diameter made of corrugated steel. for comparison, the Haverah Park tanks were 6 ft x 4 ft x 4 ft deep.}
   \label{Fig-HWAC_water_tank}
   \end{center}
   \end{figure}

Another air shower array using water Cerenkov tanks is the High Altitude Water Cerenkov Gamma-Ray Observatory\footnote{https://www.hawc-observatory.org/} (HAWC) located on the flanks of the Sierra Negra volcano near Puebla, Mexico, at an altitude of 4100 meters (13,500 feet). The water tanks (see Figure \ref{Fig-HWAC_water_tank}) are huge compared to the Haverah Park tanks but still with the same basic concept.

\section{Coming to America}
    \label{S-America}
As I was getting my PhD in Cosmic Ray Physics from Leeds University in the summer of 1964, I began to investigate possible work opportunities.  The most attractive opportunity for me in England at the time was at Harwell, the main research center of the UK Atomic Energy Authority, where they did research not just in nuclear power generation but also in cosmic rays.  John V. Jelley was doing work there on detecting Cerenkov light from extensive air showers 
\citep{jelley1958čerenkov,1961ConPh...3...45J} so it seemed like a good match to follow on from my thesis work at Haverah Park.  I remember driving down to Harwell for an interview but I guess they were not impressed by a grammar-school boy with a $2^{nd}$ Class BSc, so I was not offered a job.

I was also investigating possible positions abroad, and wrote letters to workers in cosmic ray physics and related fields that my thesis supervisor, Professor Wilson, knew or that I had heard about from their publications.  Attractive job offers came in from Mort Kaplan at the University of Rochester in the US and from Kenneth McCracken in Australia.  Without much thought, I decided to accept the US offer, thinking that I would spend a year or two there, then move on to Australia, and finish up back in England in three or four years.  I guess I had the travel bug and wanted to see the world.

Moving to America had become the popular thing for new PhDs at that time. The phenomenon was known as the ``brain drain'' and became of increasing concern in the British Parliament as more and more people took this route.  I had few qualms about joining the brain drain even though the British tax payer had paid almost all the costs of my education including for my six years at Leeds University.  I got a county scholarship to cover my undergraduate expenses and a research studentship from the Department of Scientific and Industrial Research that paid me \pounds500 per year for three years as a graduate student.  But I did not feel any responsibility to repay any of this not inconsiderable expense back to the British people.  In fact, all I wanted to do was become the best possible ``citizen of the world,'' and traveling and working abroad would certainly further that goal.  I was not sympathetic at all to the pull of nationalism or patriotism that some thought should have persuaded me to stay and work off my debts to England.

And so on a clear day in October, 1964, my parents drove me down to Southampton for the transatlantic crossing to New York (the big one) on the Queen Elizabeth, one of biggest ocean-going liners of its day.  I took along a large wooden trunk with all my worldly possessions weighing well over 100 pounds so I did not even think about flying to America.  (My father had roped up the trunk very securely but this proved to be a problem when I arrived in New York and the customs officer wanted to see inside. I had great difficulty untying and then retying the ropes.)  It was an emotional farewell as I stood on the top deck waving goodbye to my parents. Even my always emotionless father seemed sad to see me go.  But I soon got into the routine of eating and sleeping on the ship, with many opportunities to meet and talk with a wide variety of people.  I remember a very vocal older American woman at our meal table for eight people telling us all very intimate details of her life in New York. I wondered if all Americans would be like that and would expect me to be equally open about my own personal life.

I famously had two girlfriends on the ship, an English girl in the steerage class below deck where I was staying and an Australian on the upper deck where the expensive cabins were.  For some reason I was allowed to move freely between the different classes so I would spend part of the evenings below deck with my lower-class girlfriend and then later move up to spend time with my upper-class girlfriend. I was throwing off all the inhibitions of my English upbringing and beginning to explore the possibilities. Fortunately, or unfortunately, my upper-class Australian girlfriend told me after several evenings that I should not get too serious with her as she was promised to a rancher in the Outback, so that quickly ended that romance.

My lower-class girlfriend was more accommodating, and in fact, I kept up a relationship with her for over a year after the trip.  She was a trained midwife traveling to Kentucky to work with the Frontier Nursing Service that provided health care to mothers and children in poverty stricken areas of Appalachia. Since its founding in 1925, it could only get enough midwives by offering scholarships to American nurses to go to Great Britain for training in nurse-midwifery, and by recruiting British nurses who were already qualified midwives.\footnote{https://en.wikipedia.org/wiki/Frontier\_Nursing\_Service}  I spent a very enjoyable and educational week with her the following summer driving around to all her patients in her personal Land Rover that she had had shipped from England.  It turned out that she was independently wealthy and had a boyfriend back in England so that relationship did not develop into anything long term either.

\section{Rochester, NY (1964 - 1967)}
    \label{S-Rochester}
    
On arrival in New York in October 1964, I took the overnight train to Rochester, arriving there just before dawn.  This turned out to be quite traumatic since my host at the university, Graeme Duthie, did not get to the station to pick me up until several hours after I had arrived.  This would not have been a problem except that Rochester had been one of the first cities to suffer from race riots that summer of 1964, and I had read all about it with some trepidation in the English newspapers.  Thus, it was with considerable apprehension that I saw only African American men on the platform and in the waiting room at that hour of the morning.  Nothing untoward happened but I distinctly sensed the racial tension in the air.  I was later to have a girl friend, Wendy, who was very much into the civil rights movement. I remember going with her to several meetings of the Congress of Racial Equality (CORE) in Rochester but I never became an active participant myself.

Life in Rochester was good.  Although I knew not a soul, I quickly made friends.  The university campus was an ideal place for a single guy to meet people, and the Americans seemed eager to get to know me. Many were self-expressed Anglophiles and I benefited tremendously from their general love of all things English or British - most Americans did not know the difference between the two, and still do not. Many people would tell me that they loved my accent.  I would say, rather rudely, that, of course, I did not have an accent, they had the accent, I just had a dialect.  I had the good fortune to find a place to stay with a dear 80-year-old lady, Mrs.\ Thompson, who treated me like her son. I eventually shared a place with Gerry Share, a graduate student in the physics department, and Herb Fink, a student at the the Rochester Polytechnic Institute. Gerry has become a lifelong friend that I still see on a regular basis. I owe him a lot for making my transition to America a pleasant experience, and most importantly for introducing me to many eligible young ladies, including the woman who I would eventually marry (see below). He also unwittingly helped me get a job as a non-US citizen at Goddard.


  \begin{figure}    
    \includegraphics*[width=1.0\textwidth, angle = 0, 
   trim = 0 0 0 0]
   		{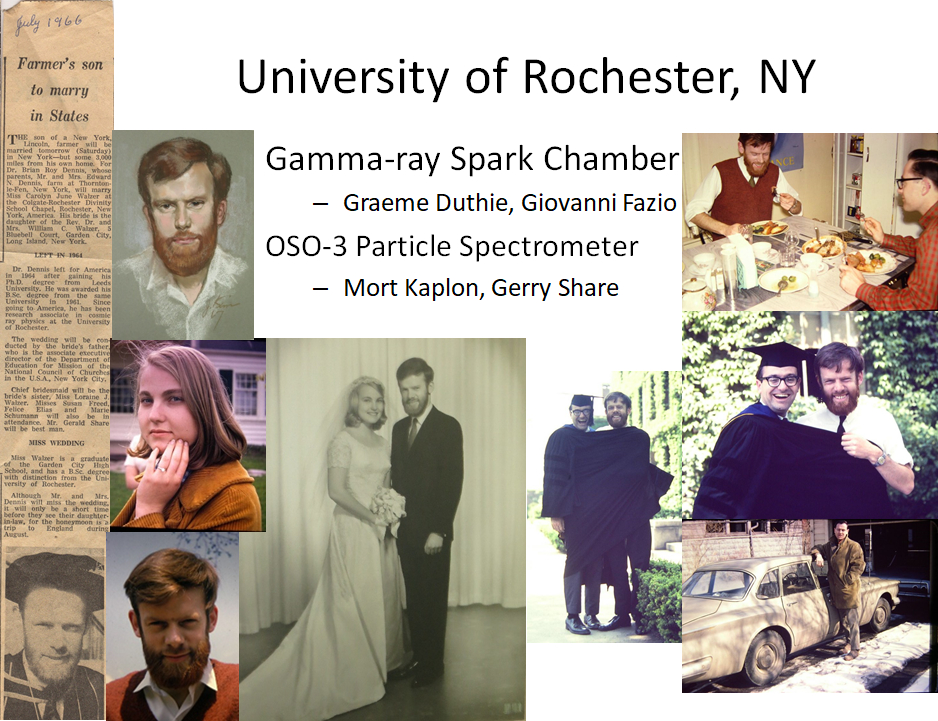}
    \caption{Pictures of me, my wife, and Gerry Share in Rochester.Top right: Dinner (that I cooked!) at the Share/Fink/Dennis apartment. Bottom right: Gerry with his beloved Zelda, a 1961 Dodge Dart.}
   \label{Fig-RochesterPhotos}
   \end{figure}

One of the first things I did when I got to Rochester was to buy an old car, a 1960's Chevrolet Bel Air. It was a police car with almost 100,000 miles on it but it had been well maintained and the price was right. I reveled in the power of this big old American tank compared to the 1939 Austin-10 I had bought for \pounds30 in 1959 and had just sold for \pounds10 before I left England.  I later exchanged the Belair (and \$1,000) for a 1964 Pontiac GTO (Gran Turismo Omologato, which is Italian meaning that it was officially certified for racing in the Grand Tourer class), the original ``muscle car'' with even more power.  It had a 389 cubic inch V8 engine, a four-barrel carburetor, and ``four on the floor,'' and was great fun to drive.  Unfortunately, like all cars driven on the salt-covered Rochester roads, it rusted out quite rapidly.  I rescued it for a while with two wooden two-by-fours holding up the gas tank, and Ken Phillips tooled around in it for a couple of years during his stay in America. I eventually sold it for \$100 to a neighbor who restored it to like-new conditions - it was that famous a model. He told me later that he would never have attempted the restoration if he had known how badly rusted out it had become.

But perhaps more important events than buying and driving fast cars happened in my two years in Rochester.  Most of them involved women.  I had established several rules to control my involvement with the opposite sex during my student days in England. The first was to not mess with women until I had got my PhD. I had been fairly successful following that rule, despite a few close calls.  At the same time, I had also developed the six-month rule, which meant that you must decide if you are going to marry somebody after dating them for six months.  No long courtships for me. This rule stood me in good stead in several instances but I finally succumbed to a blond student from Long Island who was doing a bachelor's degree in art at the University. She was younger than me in chronological years but, as often seems to be the case, much older in maturity and self confidence. Nominally, we had nothing in common since she was not at all interested in science and I ridiculed her art. Nevertheless, we seemed to get along pretty well together but after getting her bachelors degree, she moved to New York to start a masters program in elementary education at Cornell. I visited her there once during one of the hottest days ever. I remember that we went to see \emph{The Godfather} to get out of the heat. It turned out that she did not much like New York and I did not much like the idea of continuing to live as a bachelor in Rochester, so we had the somewhat delayed six-month discussion and decided that we would get married and she would move back to Rochester. We were married at the Colgate Rochester Divinity School chapel about a month after making the decision.  Her parents had been married there and her father was a Methodist minister so that greased the skids a bit. I knew this would please my mother since she was a Methodist and regularly played the organ at our local Methodist chapel back home in Thornton-le-fen. Nobody from my family came to the wedding but we went to England for our honeymoon so they all got to see the bride.

Despite the apparent arbitrary nature of our decision to get married, it has all worked out fine, and we are still together after over 50 years, many of them happy ones (to steal a joke from Al Franken). We have three children all now married with successful careers of their own. I do not really have an answer to the longevity of our marriage. Perhaps getting through the late '60s helped to cement our relationship. I personally enjoyed and benefited from the great social turmoil in those early years of our marriage, participating together in many of the activities that were popular at the time.  Things like Gestalt therapy, group therapy, massage therapy, feminism and conscience raising sessions, men's groups, re-evaluation co-counseling, conflict resolution training, and even marriage counseling and open marriage.  I learnt the difference between saying ``I think ...'' and ``I feel ...'', how to be more aware of your emotions, and not to be so cerebral all the time.  I never did learn to appreciate my wife's passion for art though, or for massage, and I never did get to like music.

Fortunately, we never got into drugs. I had decided early in life that I was not going to get involved with anything that was addictive.  This has served me well over the years. I have never smoked cigarettes, never over-indulged in alcohol (except perhaps once or twice to see how bad a hang-over can be), and never did drugs except once trying marijuana but like Bill Clinton, ``I didn't inhale it''.\footnote{http://time.com/4711887/bill-clinton-didnt-inhale-marijuana-anniversary/}

\section{Getting a New Job}
    \label{S-GettingNewJob}
    
When it came time for me to think about the future as my two-year postdoc position at the University of Rochester came to an end, I thought again of getting a job at Harwell.  I had stayed in touch with J. V. Jelley as he continued his research on Cerenkov radiation from extensive air showers.  In fact, he offered me a job to work with him and I accepted thinking that this would be a great way to return to England and stay in the field. I was then a bit surprised to get an invitation to come to New York to meet with an official committee, called the Hoff Board after its chairman Harry Hoff. I found out that this committee had been set up by the British parliament to recruit scientists who had come to the US as part of what came to be known as the ``brain drain'' and try to persuade them to return to Britain.  I could not understand why they wanted to see me because I had already decided to return and did not need any further inducements.  In any case, they offered to pay my expenses to come to New York for an interview and I went, not really knowing what the exercise was all about.  I do not remember much about the interview but I guess they must have asked me what I had been doing in America and what I expected to be doing at Harwell.  I told them about the job with Jelley and that I was looking forward to doing cosmic ray research in England again.  So it was a big shock when about a week later I got a letter signed by H. S. Hoff saying that ``the Board did not in the end see its way to recommending you for an offer of a Research Fellowship or an appointment in the Scientific Officer Class.'' So, I would not be able to work with Jelley at Harwell after all.

Looking back at this episode in my life, I realize that it may have been for the best.  If I had returned to England, my American wife would probably have been unhappy and may well have left me.  My salary would have been relatively low and we probably would not have been able to afford the several trips per year to America that she says she would have wanted to visit her parents.  At the time though, I was very disappointed and frustrated as I did not know what I had done wrong.  Of course, I suspected that again I had been rejected because of my lowly grammar school education and second-tier university degrees.  I learned later that the Hoff Boards had in fact been quite controversial and became known as ``Hoff's Circus.''  Some Members of Parliament thought that they had exceeded their intended role of recruiting brain-drain scientists and encouraging them to return to Britain.  It was thought to be outside their purview to be making decisions on particular job offers as they had done in my case.

\subsection{The Brain Drain}
    \label{Sub-BrainDrain}

In writing this memoir, I did some internet research on the ``brain drain'' and was amazed at what I found out about Harry Hoff, the chairman of the board that had interviewed me in New York.  The main sources on the internet are articles from the British \emph{Royal Society} \citep{2009NRRS..63...339B} and from British newspapers and magazines \citep{1958NS.......24H,2002TT.......24C,2010PhyW...23b...9B}.

 





   I learned that the worry about the \emph{brain drain} started in Britain after the war ended in 1945 (that would be the Second World War for my younger readers).  Many British scientists had been involved with the Manhattan Project to build the atomic bomb and had worked in America.  Many other scientists started going over to take jobs there after the war was over, and the numbers increased rapidly in the 1950's, especially after the Russians launched \textit{Sputnik}, NASA was formed, and the space race was on.  The Royal Society did a study to find out how many new PhDs moved to the US during this period, and in 1963 published a report entitled {\it Emigration of Scientists from the United Kingdom}. It contained the then shocking news that an average of 140 recent PhDs (12\% of the total UK output) had emigrated every year since 1952, most to the United States. The report also noted that in the previous five years the UK had also lost nine Royal Society Fellows.  Fred Hoyle weighed in on the ensuing debate by threatening to move to America himself if he did not get more research funding from the British government. Apparently this ploy was successful as he stayed in England. The Royal Society report had a big impact in Parliament and in the press, the fear being that if this continued it would seriously weaken Britain's capability to stay competitive. This was a time of postwar malaise in Britain with the feeling that it was losing out to America, and this report served to further stoke the already burning fires of anti-Americanism.

The term \emph{``brain drain''} was coined at this time by a reporter at the London Evening Standard and the issue became a popular political football with both political parties vowing to \emph{``do something about it''}.  The Times (of London of course) published 65 articles on the issue in the year following the report.  The House of Lords debated the issue and Lord Hailsham, the Minister for Science, made a famous speech decrying the ``Americans' parasiting of British brains,'' and promising to investigate it further.

One of the actions taken by Parliament to do something about the brain drain was to establish what became known as  the Hoff Boards.  These were groups of people set up by the Civil Service with the UK Atomic Energy Authority and the Electricity Generating Board under the chairmanship of a civil servant named Harry Sommerfield Hoff.  They were tasked with locating PhD scientists who had gone down the brain drain, as it were, and encouraging them to return and work in Britain.  Harry Hoff was thought to be well qualified to lead this effort since he had worked for the UK Atomic Energy Authority since the 1940's identifying potential young scientists and recruiting them to work at Harwell and other locations of that agency.  It was said that ``something like 30,000 scientists and engineers...passed though his hands.'' Apparently, one of them was me! The 1958 New Scientist article stated that he conducted ``an enormous number of interviews with, probably, an ever increasing shrewdness and percipience of character as well as a wide knowledge of the factors governing the requirements of scientists.'' This was presumably why the Hoff Board that interviewed me felt that part of its job was to filter the possible candidates and only recommend the most ``percipient'' for work at Harwell.  My failing may have been that I did not even know the words ``percipience'' or ``percipient,'' let alone what they mean.  What did they expect from a grammar-school boy?

But the most amazing discovery I made about this whole episode was in the obituary for William Cooper published in the British newspaper, The Telegraph \citep{2002TT.......24C}).  William Cooper was a novelist best known for the series of novels that began with {\it Scenes from Provincial Life} published in 1950.  But the most surprising thing was that his real name was Harry Hoff!!! There is no question that he is the same Harry Hoff who chaired the Hoff Boards since the obituary goes on to say that `he was at one time a civil servant responsible for trying to reverse the ``brain drain'' of scientists from Britain.'  The Telegraph obituary says that {\it Scenes from Provincial Life} ``became the prototype for Kingsley Amis' {\it Lucky Jim}, John Wain's {\it Hurry on Down}, Malcolm Bradbury's {\it Eating People is Wrong} and other irreverent, iconoclastic novels of the 1950s. {\it Scenes from Provincial Life} is a wry comedy of manners set in 1939 and constructed as a running report from the coal-face of life by Joe Lunn, a talented but frustrated science teacher at a Midlands grammar school, who enlivens an unsatisfactory existence by sleeping as often as he can with his girlfriend, Myrtle (she wants marriage; he refuses to be tied), becoming involved in skirmishes with authority and trying to establish himself as a writer.''

Hoff was born in Crewe in the county of Cheshire in western England and went to Crewe County Secondary School, presumably after passing the 11-plus exam. The same obituary states that ``by a sheer effort of will and despite the best efforts of a hostile headmaster, he won a place at Christ's College, Cambridge, to read Natural Sciences. Unlike me, he took and passed the required Latin examination. Peter Sturrock also took and passed the Latin test at the second attempt after spending ``three months studying nothing but Latin''  and made it to Cambridge \citep{2017SoPh..292..147S}. Both Harry Hoff and Peter reported that the transition from grammar school to Cambridge was not an easy one. Harry Hoff found himself overawed by the public schoolboys, who all looked ``bigger and stronger'' than he was; but he became life-long friends with C P Snow, his supervisor in physics whose claim to fame is his Rede lectures at Cambridge on \emph{The Two Cultures} (see \cite{1961PhT....14i..62S} for a review of the book that grew out of these lectures and that is now available from Amazon in the 2013 paperback reprint of the original 1959 edition). After graduating, Cooper became a science master at Alderman Newton's school in Leicester, which Snow had attended as a boy. Later, after wartime service as a squadron leader in the RAF, he joined Snow at the Civil Service Commission, interviewing young scientists and technologists for jobs.

Perhaps this is how my life would have evolved if I had been accepted into the RAF and had somehow managed to get into Oxford or Cambridge.  As it was, my future was decided by this novelist - I was to get a job with NASA and stay in America.  Perhaps the gentle reader who has persevered thus far with this memoir can decide if I would have been better off returning to England and writing novels myself.

\subsection{Moving to NASA's Goddard Space Flight Center in 1967}

I had applied for a postdoc position at Goddard at the same time that I applied for the position in England.  At that time in 1966, the NASA postdoc positions were run by the National Research Council.  As with the current NASA Postdoc Program (NPP), now run by the Universities Space Research Association (USRA)\footnote{https://npp.usra.edu/about/}, I had to write a research proposal as part of my application.  I said I wanted to make X-ray and gamma-ray observations of astrophysical sources. I think I mentioned the possible use of germanium detectors as a likely new technology that offered great promise at the time.  Given the unfortunate experience with the British recruitment committee (the Hoff Board), I was amazed to get not just one but two offers from two different scientists at Goddard to work with them.  Both Carl Fichtel in the Astrophysics Division and Ken Frost in the Solar Physics Branch had a position available, and I was interviewed by both of them on a trip down to Maryland in 1966.  I basically got to choose between the two of them and decided to go with Ken Frost for no other reason than that I liked him better.  It was a decision that would set my career trajectory for many years at Goddard.

The move down from Rochester, NY, to Greenbelt, MD, was quite traumatic.  My new wife and I rented a U-haul trailer and towed it behind our Pontiac GTO with all of our worldly possessions. My old colleague from the University of Rochester and best man at my wedding, Gerry Share, allowed us to put everything in his apartment storage area while we left and went on a month-long vacation before starting at Goddard in 1967. We were unable to get any help from NASA in finding an apartment close to Goddard since almost all the housing at that time was racially segregated, and the US government was not allowed to recommend segregated accommodation. We eventually found a convenient one bedroom apartment in one of the many so-called ``garden apartment'' complexes one stop south of Goddard on the Baltimore-Washington Parkway.

\section{Kenneth J. Frost}

I owe a great debt of gratitude to Ken Frost for taking me on initially as a postdoc for two years and then getting me a civil service position even though I was not a US citizen at the time.  (I became a US citizen in 1975 or thereabouts.) Ken was my mentor and friend and gave me complete freedom to pursue my own research. I owe it to him to include the following summary of his many contributions to NASA and to solar physics during his long and productive career.

Kenneth John Frost\footnote{https://en.wikipedia.org/wiki/Kenneth\_John\_Frost} was born in Brooklyn, New York. He graduated from Holy Trinity High School in 1952, and obtained a Bachelor of Science degree from Manhattan College and a master’s degree in physics at the University of Rochester in upstate New York. He moved to Goddard in 1958 and spent the rest of his career there until his retirement in 1997.  Sadly, he died before his time in 2013, aged 78, from complications following a fall from the roof of a backyard shed he was building at his home in Annapolis.  

Ken Frost, as he was known to all who worked with him, was a pioneer in the early space program, designing and flying instruments to detect and measure X-rays and gamma-rays in space.  He was one of the early NASA hires in 1959 at the newly opened Goddard Space Flight Center, and worked in the Solar Physics Branch at Goddard.  The Sun was his first and major concern although he was interested in any and all sources of these high energy photons, no matter where in the universe they came from.  He was the principal investigator (PI) of many flight instruments and served for several years as head of the Solar High Energy Branch.  He went on to help define Goddard's role in the early ill-fated efforts to do science on the Space Station, and became the Associate Director of Space Sciences with Steve Holt as the Director.  It is impossible to imagine anyone being allowed to fill any of these roles today without the necessary PhD qualifier that Ken never had.

Perhaps Ken’s highest honor and the one, I suspect, gave him the greatest satisfaction, was receiving the John C. Lindsay Memorial Award\footnote{https://scicolloq.gsfc.nasa.gov/Lindsay\_Awards.html} in 1982 for his role as Project Scientist and one of the prime instigators of the Solar Maximum Mission (SMM).  This award is named after John Lindsay, the man who hired Ken more than twenty years earlier and who was responsible for starting the series of Orbiting Solar Observatories (OSOs) that produced many of the advances in solar physics and astrophysics in the 1960s and '70s.  The Lindsay Prize is Goddard's highest science award given each year “to recognize the Goddard employee who best exhibits the qualities of broad scientific accomplishments in the area of Space Science.” I was deeply honored to get this award myself in 2004 thanks to a very flattering nomination written by Stuart Jordan, the head of the Solar Physics Branch at the time.

Unfortunately, it seems there is a danger that Ken Frost's legacy at Goddard and in the larger scientific world will be lost.  This is partly because Ken was not a prolific writer - only seven papers are in the refereed scientific literature with him as lead author.  Of those, only two have significant scientific results.  Again, this would not be tolerated today for a leading scientist who aspires to a successful and productive career.  The most famous paper \citep{1971ApJ...165..655F} is based on observations of a solar flare with Ken's instrument on OSO-5 and presents ``Evidence from Hard X-Rays for Two-Stage Particle Acceleration in a Solar Flare'' (see Section \ref{S-Two-stage}).  It is still cited today with a total of 154 citations with five in 2013 alone.  The second paper \citep{1969ApJ...158L.159F} and entitled, ``Rapid Fine Structure in a Burst of Hard Solar X-Rays Observed by OSO-5,'' is less well known but still has 116 citations listed.

\begin{figure}    
   \centerline{\includegraphics[width=1.0
   \textwidth,clip=]
   		{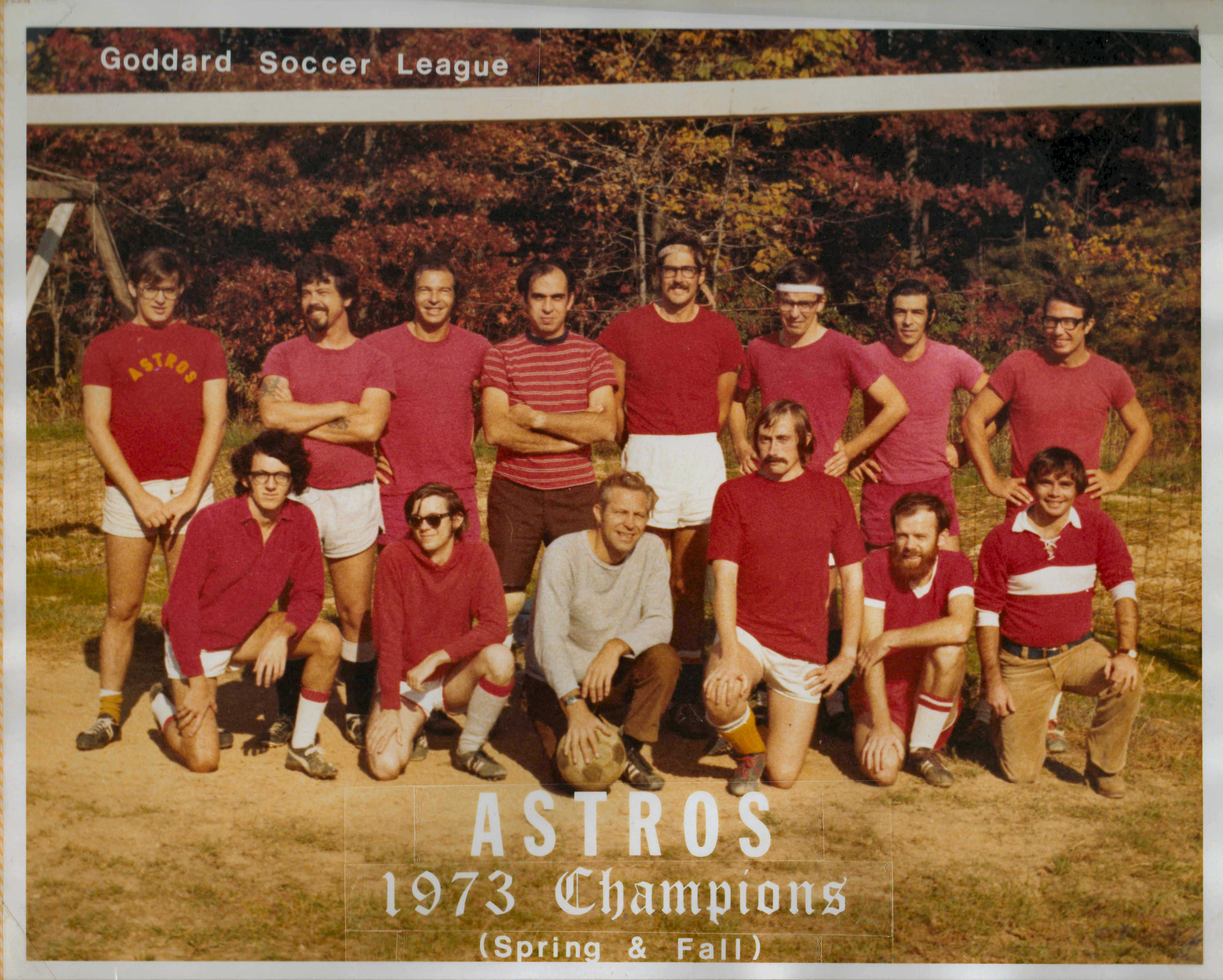}
              }
              \caption{The Astros, champions of the Goddard Soccer league in 1973. You might recognize, Pete Serlemitsos, Mario Acuna, Rick Rothschild, Peter Kenney, Art Serlemitsos, and Norm Ness standing in the back row, and Dave Thompson, Chuck Conner, Clive Dyer, myself, and Larry Orwig in the front row. Ken Frost originally played in goal but must have retired from the team by this time.         }
   \label{Fig-Astros}
   \end{figure}

It is even difficult to find evidence of Ken’s contributions in a wider internet search.  If you Google ``Ken Frost Goddard,'' you come up with just four relevant hits.  The first is to an article\footnote{http://www.goddardsoccer.com/about.php} that I wrote myself crediting Ken with a “show of interest” that led to the creation of the Goddard Soccer League in 1968. I was the captain of the Astros, one of the first four teams in the league, shown in Figure \ref{Fig-Astros} as the 1973 champions.  The league still flourishes, now with eight teams, 50 years after it was established.

The second and third hits are to brief obituaries but the fourth hit is perhaps the most interesting.  It is to an article in the April, 1966, issue of the New Yorker magazine about the failure of the \textit{Orbiting Astronomical Observatory 1} (OAO~1), NASA's first attempt to put a telescope into space\footnote{http://www.newyorker.com/magazine/1966/04/23/o-a-o}. The article states that ``The satellite failed to report in during its second day in orbit. Although the loss of the satellite (whose worth is estimated at around \$60 million, or as much as the George Washington Bridge) was a major disappointment, it would have been a miracle had the O.A.O. worked the first time, since it is such a complex piece of equipment.'' Ken was the PI of the Gamma-Ray Telescope on this mission and had high hopes that it would make the most sensitive observations ever obtained of X-rays and gamma-rays from the Sun and from anywhere else in the universe. This ill-fated spacecraft was launched on 8 April 1966 but high-voltage arcing in the star trackers and a failure in the power supply system caused the mission to be terminated after 20 orbits without activating any of the experiments.  This setback must have been devastating to a young scientist who had put his all into developing the largest and most sensitive X-ray and gamma-ray spectrometer of its day.  Nevertheless, Ken and his team carried on, as reported in the New Yorker article that concludes with the statement – ``After the experiment failed, Dr.[sic] Kenneth Frost, and another Goddard astronomer, Henry Horstmann[sic], were making plans to send the Goddard experiment up in a balloon, as they had originally planned to do.'' They proceeded to do just that from Palestine, Texas, with several successful flights in the following years, some of which I participated in.

In fact, Ken Frost was the PI for many other flight instruments that made lasting contributions to high energy solar physics and astrophysics.  He was PI for instruments on four of the eight \textit{Orbiting Solar Observatories} – OSO-1, -2, -5, and -8.  All the instruments were X-ray/gamma-ray spectrometers of the same basic design that had been developed for use in the laboratory on the ground.  Ken's task was to make these very fragile devices survive the intense vibrations during launch and operate successfully in the vacuum of space for long periods of time. 
His competitors were Larry Peterson at the University of California, San Diego (UCSD), Ed Chupp at the University of New Hampshire (UNH), and others. They developed similar detectors that were flown on OSO-1, -3, and -7 so that there was an X-ray/gamma-ray spectrometer on almost every one of the eight OSOs.  This emphasis on X-ray and gamma-ray observations was in recognition that these high-energy photons, not detectable from the ground, offered a new window on the universe.  They had the potential for enabling new scientific discoveries that could rival or exceed all that had been achieved previously with ground-based optical astronomy.  Hence, there was a great demand for flying ever-improving X-ray and gamma-ray instruments, and the OSO series offered many opportunities to do that for almost twenty years.

Ken Frost's contribution to the early days of solar physics observations from space should not be forgotten despite his few published science papers.  There are, of course, many other papers in the literature with important science results based on the analysis of data from Ken's instruments but his role is not always made clear.  Perhaps his greatest contribution, though, was to the whole political process of devising and advocating for new instruments and new missions.  This is the aspect of his work that always seemed to interest and motivate him the most.  And he was a master at it as evidenced by his success in getting so many instruments on the different OSO spacecraft.  But he will be best remembered for his role in instigating and ushering in the \textit{Solar Maximum Mission}.  While it did not achieve the goal he had hoped for of revealing a complete understanding of solar flares, it did set the foundation on which all future solar space missions depend.  More people are needed today with Ken's vision, and his ability to navigate the political landscape for the betterment and advancement of solar physics and astrophysics.

\subsection{Active Anticoincident Shields}
\label{sc_anticoincidence}

My first job at Goddard was learning about actively shielded X-ray detectors that Ken Frost had designed and built for both astrophysical and solar observations.  Ken was in the Solar Physics Division at Goddard but was still allowed to make non-solar observations even though that was really the purview of people in the Astrophysics Division led at that time by Frank McDonald\footnote{https://en.wikipedia.org/wiki/Frank\_B.\_McDonald}.  This separation led to some tension but Ken didn't want to be under Frank's thumb and preferred the relative freedom to do what he wanted under Jack Brandt, the head of the Solar Physics Division. Ken was always keenly interested in politics, both local and national, and would delight in arguing both the liberal and the conservative side of every issue depending on who he was talking to at the time. Like a true debater, you never really knew which side he really believed but you always learned a lot from his insight. He became a past master at using the NASA system to his own advantage, and I think certainly to the advantage of solar physics.

In any case, Ken was known for being the first person to use an active anticoincidence shield to reduce the background rate in X-ray detectors flown in space.\footnote{https://en.wikipedia.org/wiki/Electronic\_anticoincidence}
This design arose after initial flights of bare X-ray detectors on balloons had revealed a large background counting rate that was attributed to high energy charged particles from all directions passing through the detector and making it difficult to see the X-ray and gamma-ray photons of interest from the Sun or from more distant astrophysical sources. 

The first X-ray detection of a flare was from a balloon-borne instrument with an integrating ionization chamber and a single Geiger counter \citep{1959JGR....64..697P}. In order to increase the sensitivity and provide some directional information, the two instruments on the first \textit{Orbiting Solar Observatory} (OSO-1) were passively shielded scintillation detectors.  Ken Frost's instrument was a cylindrical NaI(Tl) crystal 0.3~cm high and 2.24~cm in diameter placed inside a copper shield with a wall thickness of 1~cm \citep{1962ITNS....9..381F}. It recorded the rate of pulses from the photomultiplier tube (PMT) viewing the scintillator with amplitudes in the range corresponding to energy losses in the crystal between 20 and 100 keV.

The early history of this work has been described by Larry Peterson \citep{2013AAS...22111309P,2014...P}.  He tells how he met Ken Frost in $\sim$1962 after flying his gamma-ray experiment on OSO-1. Peterson writes that ``During a meeting with Lindsay at GSFC, I found that a member of his group, Kenneth Frost, was already developing a NaI counter with an active CsI(Tl) collimating shield for the second OSO. After a discussion of problems in implementing such a device, Ken and I formed a collaboration to do a balloon test of the proposed configuration. The test flight was flown from Minneapolis in the summer of 1962. Besides obtaining limits on hard X-ray emissions from the quiet Sun, extensive data on atmospheric $\gamma$-rays, instrumental background, phosphorescence decay effects, and anti-coincidence requirements were obtained \citep{1966JGR....71.4079F}). Knowledge from that experiment was incorporated into later GSFC and UCSD instrument designs.''

The first attempt to reduce the background involved using a thick lead shield around the X-ray detector with a hole in it to allow photons to enter from the direction of interest. Unfortunately, this only made things worse because of all the secondary particles produced when the charged particles interacted with the lead.  The solution was to use a particle detector as the shield and to electronically reject all events recorded by the central detector if there were simultaneous signals from the shield. 

\begin{figure}    
   \centerline{\includegraphics[width=1.0
   \textwidth,clip=]
   		{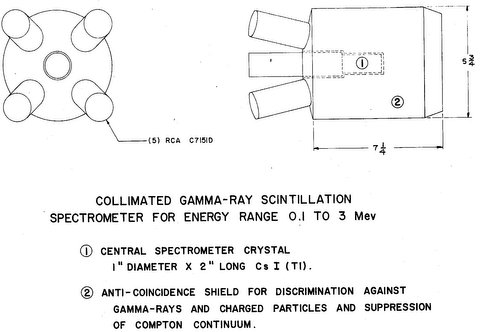}
              }
              \caption{Ken Frost's first active anticoincidence collimated scintillation spectrometer designed for gamma-ray astronomy in the energy range from 0.1 to 3 MeV \citep{1962ITNS....9..381F}.         }
   \label{Fig-antico}
   \end{figure}

For his first gamma-ray spectrometer, Ken used the same inorganic scintillator for both the shield and the central detector as shown in the Figure \ref{Fig-antico}.  He chose cesium iodide doped with thallium, called thallium activated cesium iodide, or CsI(Tl). These inorganic crystals give off a burst of scintillation light when traversed by a charged particle, with the intensity of the light being very close to linearly dependent on the energy deposited in the crystal.  They also scintillate when an X-ray photon interacts inside to produce a charged particle as, for example, through the photoelectric effect, Compton scattering, or pair production. At the time, they were the best X-ray and gamma-ray detectors available and allowed the energy spectrum of the incident photon flux to be determined over a wide range of energies from $<$10~keV to $>$10~MeV.  The energy resolution was not great, being $\sim$50\% at 10~keV improving to $\sim$20\% at 1 MeV.  This was not a serious problem since the measured spectra of solar flares and cosmic sources in this energy range are generally smoothly varying with the photon flux decreasing with increasing energy according to a power-law like function. Few sharp features are present in this energy range that require fine energy resolution.

The trick that Ken used to minimize the background counts was to only record a signal from the central X-ray detector when there was no simultaneous signal from the shield.  This is called operating the shield in anticoincidence with the central detector. In this way, all charged particles are rejected if they produce a simultaneous signal in both the shield and the central detector.  

\begin{figure}    
    \includegraphics*[width=1.0\textwidth, angle = 0, 
   trim = 0 15 0 0]
   		{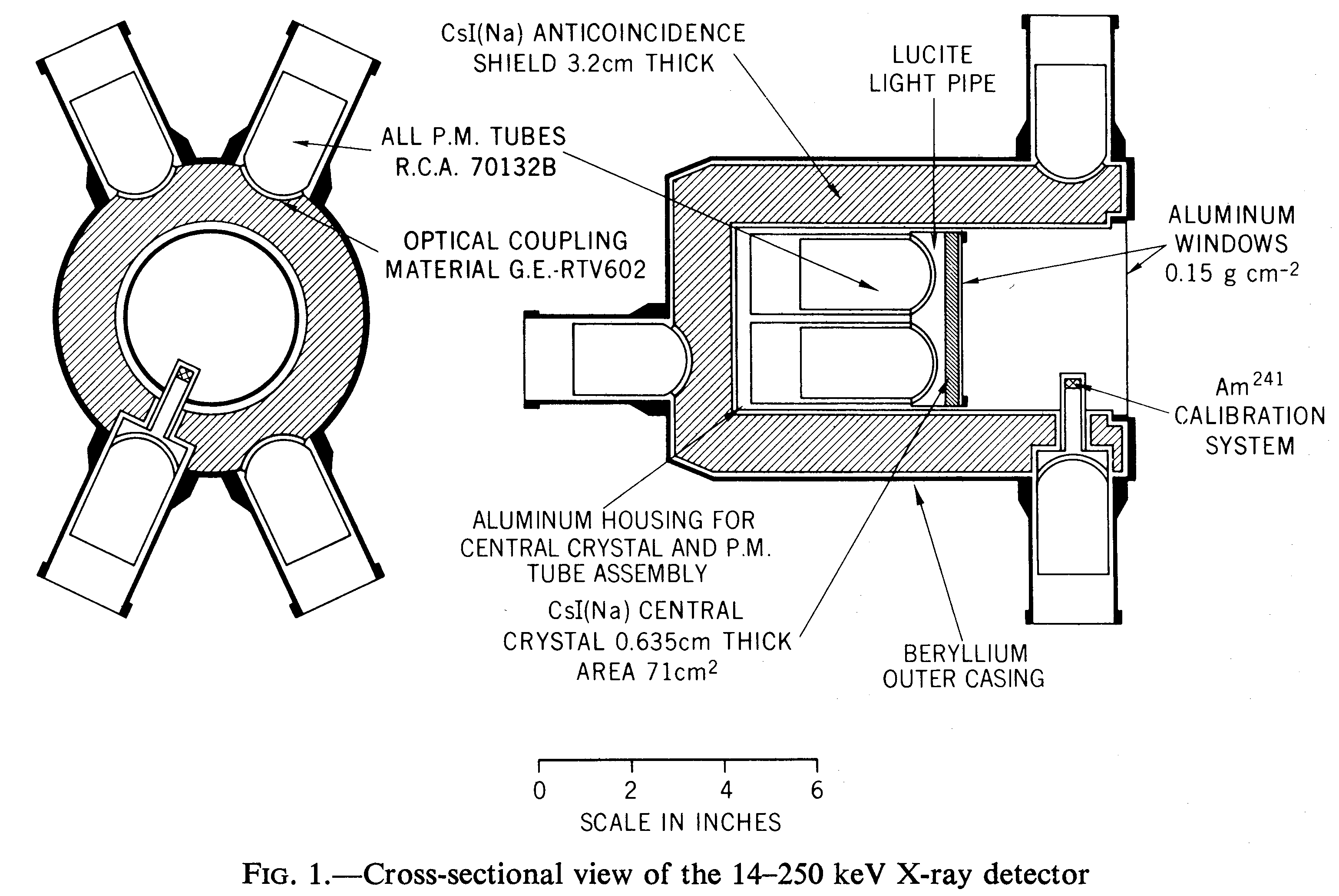}
    \caption{Cross-sectional view of the X-ray Spectrometer flown on OSO-5.}
   \label{Fig-OSO5Spectrometer}
   \end{figure}

According to \cite{1969ARAA...7..121N}, active anticoincidence reduced the background by a factor of ten over the background encountered in the absence of active shielding. The method was first used in space by \cite{1962ITNS....9..381F} on OSO II and by \cite{1966JGR....71.4079F} on OSO III for the detection of solar and celestial X-ray sources. The modified design used by Ken Frost for the X-ray spectrometer on \hbox{OSO-5} shown in Figure \ref{Fig-OSO5Spectrometer} was optimized for observations of X-rays between $\sim$20 and 250~keV from solar flares.

This basic active anticoincidence design has been used for most X-ray and gamma-ray spectrometers for observations of both solar flares and astrophysical objects. Only with RHESSI was a separate active shield dispensed with because of weight limitations but also because Bob Lin realized that for the detection of transient events like solar flares, it is more important to have the largest possible sensitive area to detect the most photons than to reduce the detector background level down to the lowest possible value. Even with RHESSI though, some effective active shielding is provided by electronically dividing each germanium detector into two segments, a thinner front segment and a thicker rear segment.  The front segments shield the rear segments from the intense fluxes of low energy photons during flares while the rear segments serve as active shields for the front segments.

\subsection{Pulse Pile-up}

Perhaps the single most important aspect of the design that led to the considerable success of the OSO-5 instrument, and similar instruments on OSO-8, and SMM, was the inclusion of relatively thick aluminum windows in front of the central detector.  These windows were included specifically to absorb the lower energy incident photons, especially those below 20 keV that had plagued the earlier X-ray spectrometer from UCSD flown on OSO-3 as detailed by \cite{1970SoPh...14..414K}. That detector had only a thin beryllium window (93~mg~cm$^{-2}$) with 50\% transmission at 4.1~keV. Such low energy photons can become so intense during even modest solar flares that they cannot be handled by the PMTs and the associated electronics. The specific problem arises when the counting rate becomes sufficiently high that there is a significant probability that two or more photons will arrive during the e-folding decay time of the scintillation light, $\sim$2~$\mu{s}$ for CsI(Na). In that case the resulting spectrum of pulse amplitudes becomes seriously distorted. 

Attempts to correct for this effect and recover the spectrum of the incident X-ray flux have met with limited success. The first attempt was made by \cite{1975SSI.....1..389D}.  A computational technique was developed to simulate the effects of pulse pile-up and use it to model the UCSD solar hard X-ray experiment on OSO-7.  \it ``Excellent agreement'' \rm was claimed between the predictions of the model and the observations of three large flares but nevertheless it was recommended that \it ``thick entrance windows and short resolving times in the system electronics'' \rm be used to minimize pile-up. This advice is still true today because no system has been devised to accurately correct for pulse pile-up, at least not for X-ray detectors when the probability of two pulses occurring within the electronics resolving time is $\geqslant$10\%.

Few instruments have been flown capable of handling the full dynamic range of solar flares from the smallest GOES A1 class events to the largest X20 events, a range of over five orders of magnitude.  Perhaps the best attempt was made with the movable attenuators implemented on RHESSI \citep{2002SoPh..210...33S}.  One or two sets of thin aluminum disks were automatically moved in front of each of the nine germanium detectors of this instrument when the count rate exceeded adjustable threshold levels.  In this way, the count rate can be kept below the level at which pulse pile-up is significant, except for the largest events.  

\section{Science with Ken Frost}

Ken hated having to write papers. He claimed, like Peter Sellers in the movie \emph{\mbox{Being There}}, that he didn't know how to write.  He was very good at explaining things orally but his handwriting was terrible and he was very bad at forcing himself to sit down and write up some observation or idea that he was working on. So he was not the most productive person in terms of the number of scientific papers that he published.  Nevertheless, he did write a couple of seminal papers that are still frequently cited to this day. 

\subsection{Two-stage Particle Acceleration}
	\label{S-Two-stage}

\begin{figure}
  \begin{center}
    \includegraphics*[width=1.0\textwidth, angle = 0, 
   trim = 0 0 0 0]
   		{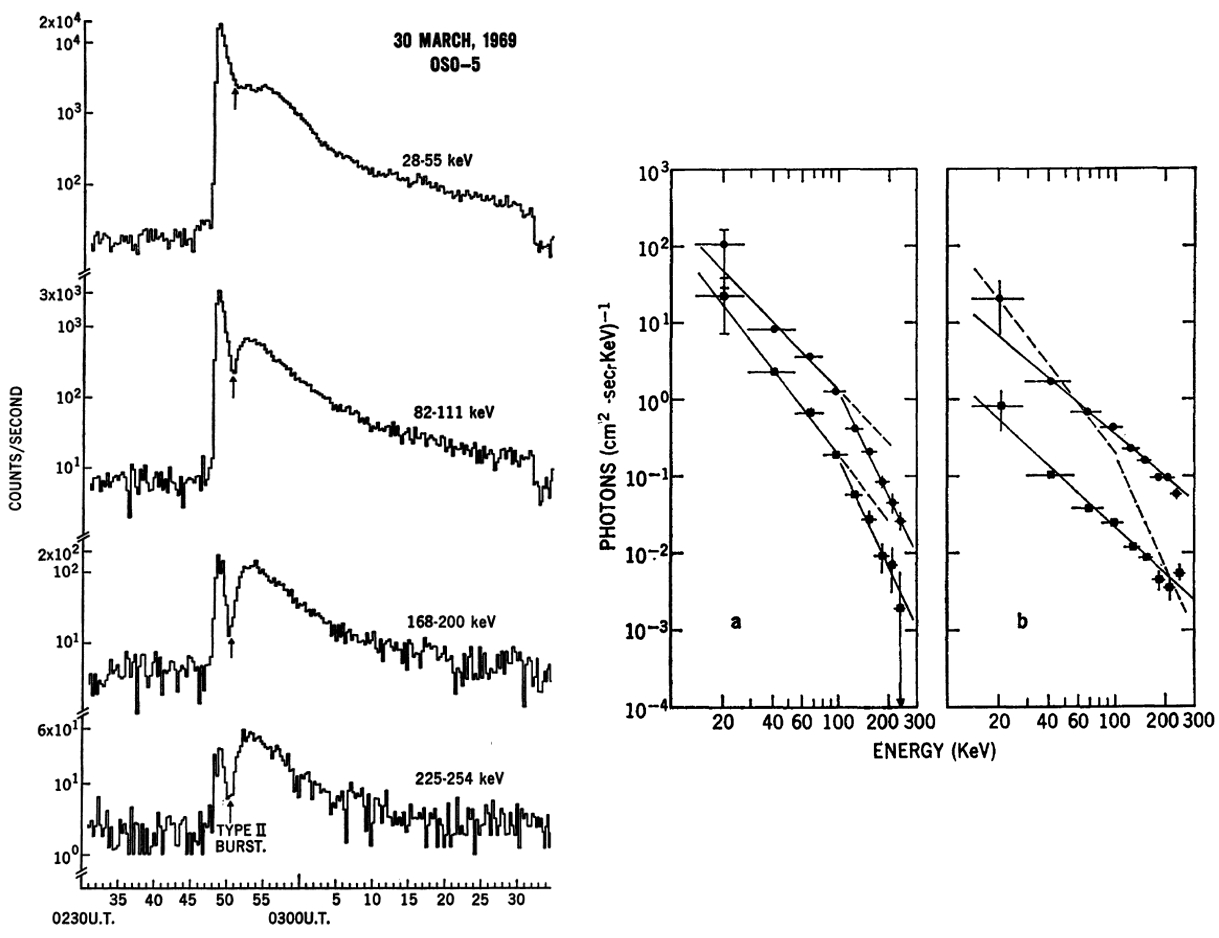}
    \caption{\textit{Left}: X-ray count rate time profiles of the flare on 30 March, 1969, in four different energy bands showing evidence of two stages of electron acceleration. The vertical arrow shows the start time of the radio Type II burst. \textit{Right}: (a) Broken power-law X-ray spectra during first peak. (b) Flatter power-law spectra during the second peak. (from \cite{1971ApJ...165..655F})}
   \label{Fig-FrostDennis}
   \end{center}
\end{figure}

Perhaps his best known paper is \cite{1971ApJ...165..655F} on the analysis of X-ray observations of a solar flare made with Ken's instrument on OSO-5. I don't remember exactly what my role was to become a co-author but I suspect that, since Ken was not too good with computers, I wrote the Fortran programs used to analyze the data and produce the plots in the paper. Figure \ref{Fig-FrostDennis} shows the X-ray light curve from the paper with the two distinct phases, a short impulsive peak followed by a more gradually varying peak.  At first glance, it doesn’t look too remarkable but Ken had become familiar with the concept of two stages of electron acceleration in some large flares that had been proposed by Australian radio astronomers based on their microwave observations 
\citep{1963ARA&A...1..291W}.  This was the first published X-ray evidence that supported this hypothesis.  I think Ken had met them at one of the few international scientific meetings that he attended, this one I think in a place called Surfer's Paradise on the east coast of Australia. 

In this two-stage model, the electrons accelerated in the first stage up to \mbox{$\sim100$~keV} produce the first peak in the X-ray light curve and electrons accelerated in the second stage to higher energies and with a harder spectrum produce the second, more gradually varying peak. 
\cite{1969sfsr.conf....1D} had proposed that the first stage acceleration was through an induced electric field, perhaps arising in a “pinch-type” instability, and that the second-stage was through a Fermi mechanism in a shock front.

Hugh Hudson has started calling these types of two-stage flares Frost-Dennis events \citep{2017arXiv171105583H}. In fairness, I think flares with the two-stage X-ray signature could be called Frost events but the two-stage model itself should be called the Wild, Smerd, Weiss, and de Jager (WSWD) model.

These two-stage events have recently received renewed attention with the observations of high energy gamma-rays above 100 MeV using the \textit{Large Area Telescope} (LAT) on the \textit{Fermi} spacecraft.  Many solar events have now been detected at these energies that show two stages, an initial impulsive stage lasting for minutes followed by a generally much longer stage lasting for up to 20 hours (e.g.~\cite{2017arXiv171101511S}). A likely explanation for these events is based on the familiar two-stage particle acceleration model but now applied to both electrons and protons.  The LAT gamma-ray spectra of these events is best fit with that expected from the decay of pions produced by $>300~MeV$ protons interacting with ambient protons in the solar atmosphere most probably in the photosphere, where the density is high enough to produce the observed gamma-ray flux.  One model proposed by \cite{2017arXiv171101511S} has protons accelerated to \mbox{$\sim1$~MeV} in the first phase.  Some of them traveling at a few thousand $km~s^{-1}$ can catch up with the shock produced in front of the associated CME traveling generally at less than 3000 $km~s^{-1}$. These protons can become seed particles for further acceleration to GeV energies in the shock.  Those accelerated outwards onto open magnetic field lines are the solar energetic particles (SEPs) detected in space; those accelerated back towards the Sun could eventually reach a region of sufficiently high density in the photosphere and produce the observed $>100~MeV$ gamma-rays. 
\cite{2017arXiv171105583H} has proposed what he calls a magnetic lasso to overcome the mirroring effect of the converging magnetic field closer to the Sun, which would prevent all but a very small fraction of the shock accelerated protons from reaching the photosphere.

\subsection{Diffuse Cosmic X-ray Background}

\begin{figure}
	\begin{center}
    \includegraphics*[width=1.0\textwidth, 
    angle = 0, 
   trim = 0 0 0 0]
   		{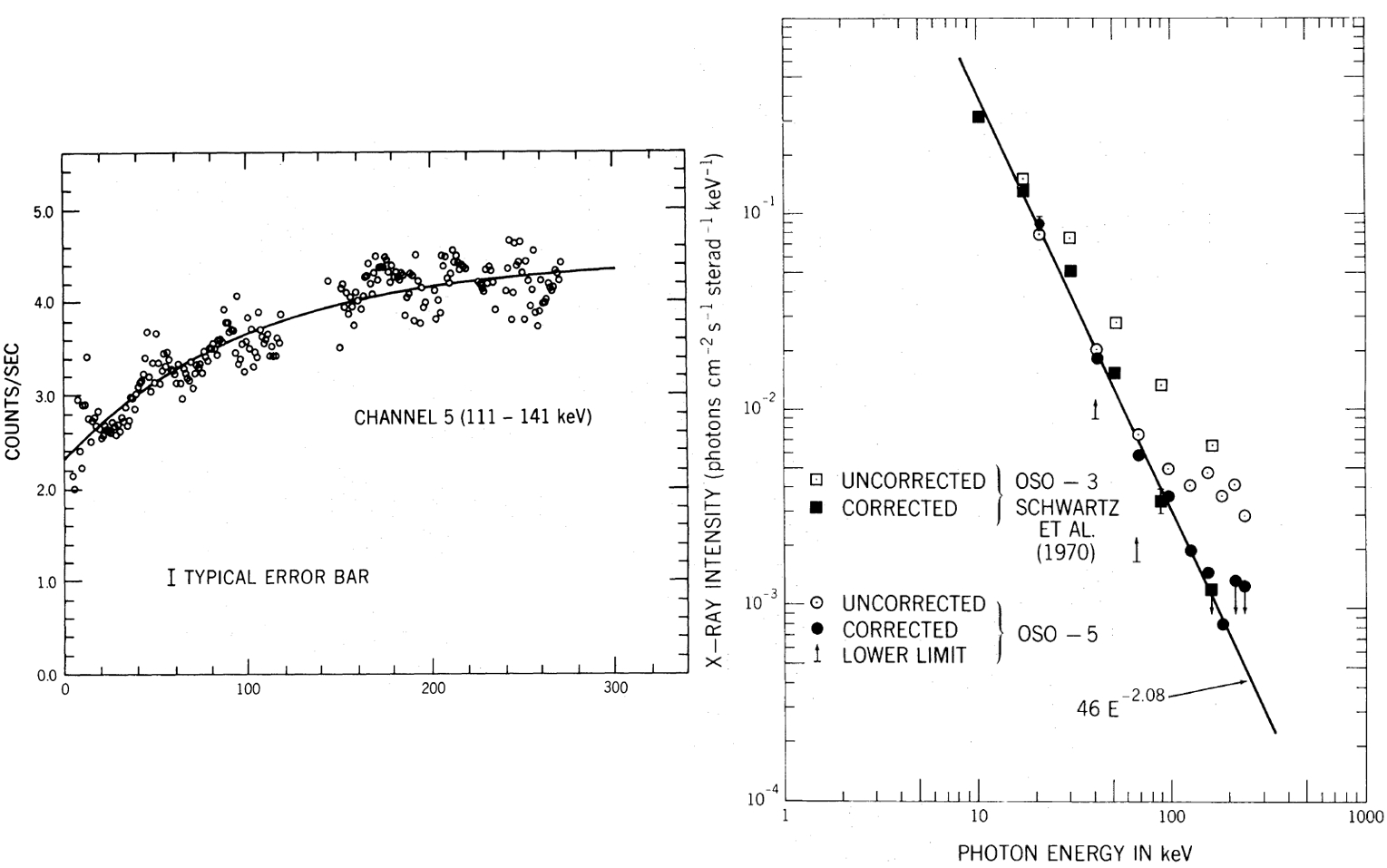}
    \caption{\textit{Left}: Buildup in the 111--141 keV background counting rate of Ken Frost's instrument on OSO-5 \textit{vs.} days after launch. \textit{Right}: Spectra of the diffuse cosmic X-ray background measured with instruments on OSO-3 and OSO-5. (from \cite{1973ApJ...186...97D})}
   \label{Fig-DCBspectrum}
   \end{center}
\end{figure}

Apart from my involvement with the \cite{1971ApJ...165..655F}
paper, I concentrated on the non-solar aspects of the OSO-5 observations, notably the spectral measurements of the diffuse cosmic background radiation. The $\sim40^{\circ}$ FWHM field of view of the X-ray spectrometer (see Figure \ref{Fig-OSO5Spectrometer}) made it a useful instrument to detect and measure this largely isotropic radiation that has unknown origins and was the subject of considerable scientific interest.  Since the flux of this emission is essentially constant in time, the detector background could not be so easily determined as it could for measuring the flux from transient events like solar flares.  After much effort to remove the orbitally varying background, I finally published my best estimate of the diffuse cosmic background spectrum 
\citep{1973ApJ...186...97D}, my first significant paper as lead author.  This was not the first measurement of the hard X-ray spectrum but it was comparable to the earlier results from the University of California group in San Diego (UCSD) 
\citep{1970ApJ...162..431S}. Perhaps it didn't contribute that much to understanding the origin of the diffuse emission but I couldn't understand why my paper is not cited more often in later reviews of this field. 
Not until I discovered that the on-line electronic version in NASA ADS is missing the crucial last three pages with the main results. Apparently, they had been missed when the paper was scanned.  This has recently been corrected thanks to my bringing it to their attention but, in any case, here is the plot in Figure \ref{Fig-DCBspectrum} of the spectrum showing the remarkable agreement with the UCSD spectrum after significant corrections were applied in both cases.

My result is included as one of 45 published spectra plotted in Figure 1 of the excellent review of measurements of the X- and gamma-ray diffuse background by \cite{1975NCimR...5..255H} and copied in the more recent review by \cite{2017SSRv..212..429C}. The 1975 paper is by my good friend Henry Horstman, who worked at Goddard in Ken Frost's group in the 1960s. He initiated many happy spelunking trips exploring the caves in West Virgina with other Goddard colleagues including Ed Rothe and Charles Wolff. Henry was a confirmed, chain-smoking bachelor at that time until he met and married the Italian scientist, Elena Moretti, and moved to Bologna with her for the rest of his productive but too-short life.

\subsection{Review Papers}

Other first-authored papers that I am more pleased with and that are more frequently cited are my review papers.  The first was by invitation to write a review of solar X-ray bursts for the \(100^{th}\) volume of Solar Physics 
\citep{1985SoPh..100..465D}.  I accepted the task somewhat reluctantly as by this time I had caught Ken Frost's dislike of writing papers. I would much rather do something new and much more interesting instead of rehashing what I had already worked on.  In any case, I embarked on writing this review paper and I actually found it to be very rewarding. Writing about other people's work forced me to read more papers on the subject.  Also, many people in the field were very helpful in explaining their work to me at meetings and during their visits to Goddard.  For the almost ten-year life of SMM in the 1980s, many people came to Goddard and stayed for extended periods.  This was before it was possible to easily participate remotely in the daily planning meetings or to transfer data to your home computer.  Consequently, most of the PI teams were based at Goddard, and I got to know many of the most productive people in the field during those years.  

The process of writing a review paper and preparing it for publication was quite laborious and time consuming at that time.  Papers and references could only be found in the library. I generated an extensive library of my own of Xeroxed copies that I have to this day in four drawers of a filing cabinet in my office. Perhaps I will eventually shred them all but not yet. What did people do before Xerography was invented?  I would write each section in my barely legible hand writing (but better than Ken Frost's) and give it to the secretary for typing.  There would inevitably be many typographical errors in the first draft that I had to mark up and give back for retyping.

Despite this now antiquated process, the resulting paper was quite satisfying, and I learnt (learned?) a tremendous amount of solar physics in the process.  I would strongly recommend that any postdoc relatively new to a field should write a review paper.  While there is some truth to the somewaht modified old adage that \textit{those who can, do; and those who can't, write review papers}, it is a great way to broaden your knowledge and further your interactions with other people worldwide.  

People seemed to like my review and it has often been cited as a general introduction to early solar X-ray observations. So, fortunately or unfortunately, I become branded as a writer of reviews and I was asked to write further papers of that sort. Unfortunately, the process got more and more tedious to the point that I just couldn't finish the last one that I had committed to write.  I am eternally grateful to Richard Schwartz for helping me complete it and get it published as \cite{1989SoPh..121...75D}. 

While I did write a shorter review of the early RHESSI results with help from Hugh Hudson and S\"{a}m Krucker \citep{2007LNP...725...33D}, I have avoided doing anything similar until I embarked on this memoir. I realize how much easier it is now, with everything available on line including computer graphics, NASA ADS, Latex, BibTeX and Overleaf, and almost instant access to almost any paper ever published. Unfortunately, Goddard is no longer the center of solar physics as it was during the SMM era in the 1980s.  While most solar physicists still visit Goddard on occasion, few of them stay for extended periods, and I miss the long-term interactions with them.  Also, I still find myself more comfortable handwriting the first draft of this memoir. The difference now is that we don't have secretaries that will type it all up for us so I have had to learn that skill too.

\subsection{The Flare Size Spectrum}

\begin{figure}
	\begin{center}
    \includegraphics*[width=0.8\textwidth, 
    angle = 0, 
   trim = 0 0 0 0]
   		{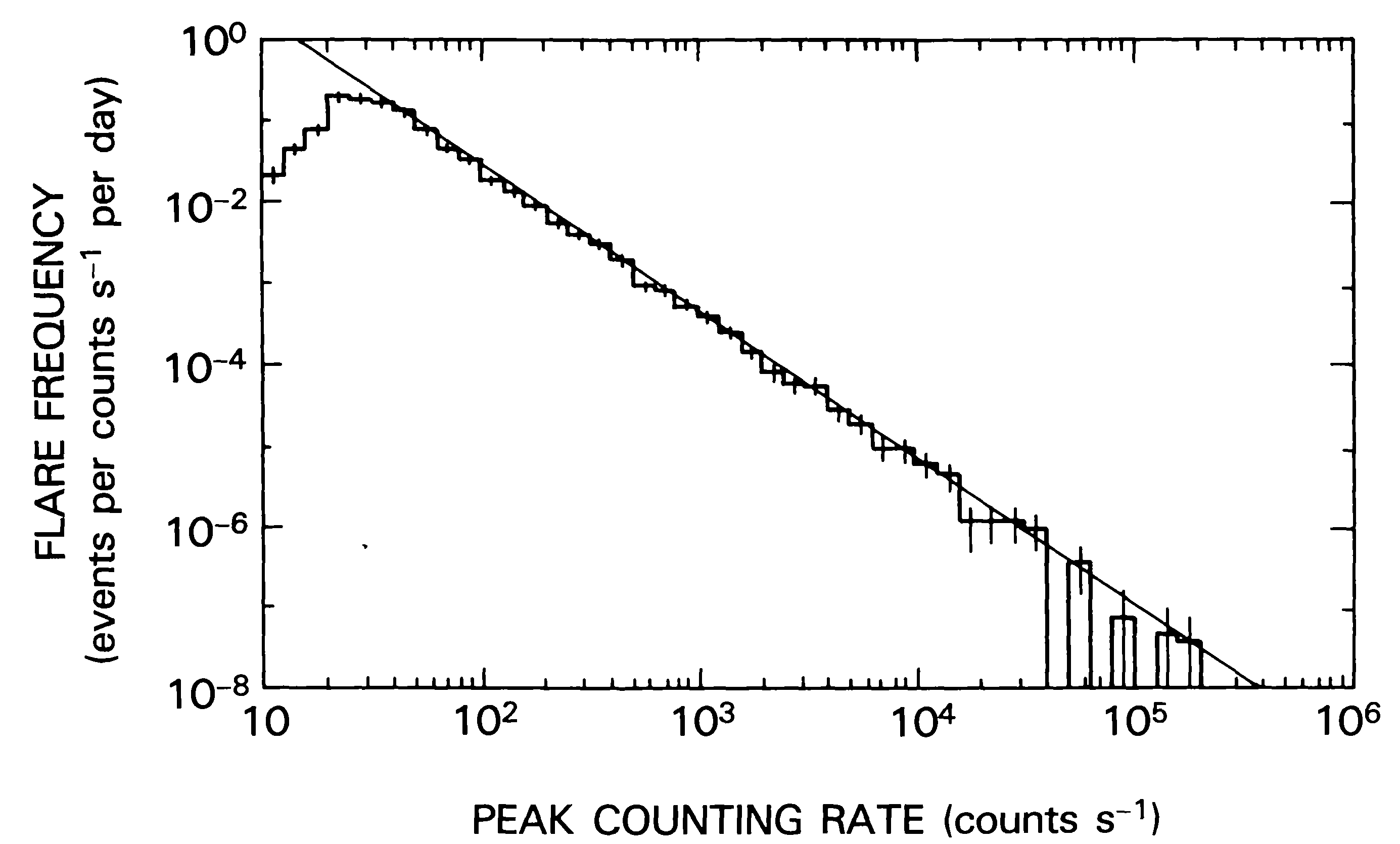}
    \caption{Peak rate spectrum of all complete events detected with HXRBS from 1980 to February 1985 (from \cite{1985SoPh..100..465D})}
   \label{Fig-FlareSizeDistributon}
   \end{center}
\end{figure}

Another positive result of writing review papers is that you can sneak in some of your own work with perhaps a little less stringent refereeing. I included the flare size spectrum that I obtained from the \textit{Hard X-ray Burst Spectrometer} (HXRBS) on SMM in \cite{1985SoPh..100..465D}. That spectrum in Figure \ref{Fig-FlareSizeDistributon} shows the number of flares detected with HXRBS with a given peak counting rate plotted as a function of the peak rate.  

This plot was not the first hard X-ray flare size spectrum ever published.  That honor goes to \cite{1974SoPh...39..155D}, who published the integral size distribution for 123 flares observed with their solar X-ray experiment on OSO-7. Their power-law index of -0.8 is consistent with my value of -1.8 for the differential distribution that I was able to use with the over 6,000 flares in the SMM/HXRBS data set. But little did I know when I made this plot that it would be so pivotal in helping people smarter than myself understand a fundamental law of nature

My result in Figure \ref{Fig-FlareSizeDistributon} attracted the attention of Per Bak, who was developing his ideas on self-organized criticality at that time \citep{1987PhRvL..59..381B}.  He knew that many phenomena, notably earthquakes and sand piles, exhibit power-law distributions of intensity, and felt that the solar flare size distribution was also a prime example. He included my plot in his book \emph{How nature works : the science of self-organized criticality} \citep{1996hnw..book.....B}. (He got the satellite and instrument wrong in the figure caption - ISEE3/ICE instead of SMM/HXRBS but that's OK.) He gave a colloquium at Goddard a few years later and I was surprised to see him show my plot during his presentation. 

Markus Aschwanden included my plot (Figure \ref{Fig-FlareSizeDistributon} in his massive book, \emph{Self Organized Criticality in Astrophysics : The Statistics of Nonlinear Processes in the Universe} \citep{2011soca.book.....A} (How does he do it???)  He tells the story in the Preface of how initially ``no explanation could be found'' for what he called my ``perfect power-law distribution" extending over almost four orders of magnitude with a slope of -1.8. 

My plot also interested Ed Lu when he was developing the avalanche model of solar flares with Russ Hamilton \citep{1991ApJ...380L..89L}. They included it in their paper and also new plots including all of the HXRBS data broken down by years during the almost complete solar cycle from 1980 through 1989 that SMM was operational. (I apparently provided these plots ``expeditiously'' according to the acknowledgements.)  They explained the observed power-law dependence of solar-flare-occurrence rate on flare size (which they claimed extended over more than five orders of magnitude in peak flux) by first assuming that the solar coronal magnetic field is initially in a self-organized critical state. They postulated that a solar flare is made up of multiple  small reconnection events, analogous to an avalanche in a sand pile. They used a simple lattice-reconnection model and showed that it produced flare size statistics consistent with the measured power-law distribution.

One thing that must be considered in addition to the variations in the flare energy release is the distribution of active region properties \citep{1995ApJ...446L.109L}. Clearly, there must be an upper limit to the flare energy from any given active region equal to the available free magnetic energy in that active region.  Thus, the measured flare size distribution must be the accumulation of multiple distributions for flares from different active regions, each with a maximum cutoff energy. Following Ed Lu's suggestion, we looked for evidence of a cutoff in the overall flare size spectrum but the small number of large flares in our sample made the results ambiguous \citep{1997ApJ...475..338K}. We did, however, find that active regions with small sunspot areas did have the expected upper limit to the energy of the flares that they produce. This must be taken into account in any attempt to explain the production of the power-law flare size distribution.

Since those early days, much work has been done on self-organized criticality. \cite{2014PhR...536...41M} point out that the ``current experimental evidence is still inconclusive with respect to a possible causal relation of the emergent power laws to an underlying self-organized critical state.'' However, the avalanche model of solar flares is still a viable possibility although \cite{1995A&A...299..897V} pointed out that since particles and plasma in the corona are constrained to flow along the magnetic field, anisotropic 1D transport should be considered rather than the 3D model used by \cite{1991ApJ...380L..89L}. 

The work on SOC is extensively reviewed and brought up to date by \cite{2011soca.book.....A}, by \cite{2014PhR...536...41M}, and by the papers resulting from the two workshops on “Self-Organized Criticality and Turbulence” at the International Space Science Institute (ISSI) in Bern, Switzerland, in October 2012 and September 2013 \citep{2016SSRv..198..167S,2016SSRv..198...47A}. The belief is still, as stated by \cite{2016SSRv..198..167S}, that ``Solar flares are one of the most compelling examples of SOC type behavior; their power-law distribution covers over eight orders of magnitude (Aschwanden et al. 2014).'' However, a detailed or even a perfunctory review of this field is well outside the scope of this paper, and certainly beyond me, given my limited understanding of this complex subject. I will note, however, that \cite{2017PhRvE..95b3209K} have studied energy release in a 2D current sheet and find that the dynamics of the system are consistent with SOC. Perhaps this aspect can be further explored for the current sheets that develop during a solar eruptive event between the rising CME and the reconnected flare loops.

\subsection{Open Data Access}

When I started work at Goddard in 1967, open data access was not an issue. It was clear that the PI responsible for a given instrument had full control of the observations made with that instrument and could limit access to the data in virtually any way they chose. Nominally, according to NASA's rules, the PI had exclusive data rights for only one year after the observations were made. In practice, however, almost all PIs retained full control indefinitely even though the tax payer had paid all the costs. I remember Carl Fichtel, PI for the EGRET gamma-ray instrument, arguing that he would not be able to get the best scientists to work on future instruments if they knew that the data would ever be available to anybody else without the consent of the PI. He argued that only PI team members were knowledgeable enough about the instrument and the critical details that were needed to correctly and accurately analyze the data. False scientific claims might be made by any other researcher since they would not be so familiar with all the idiosyncrasies of the instrument and of the data analysis routines. This closed philosophy was almost universal prior to the 1980s. Despite having similar X-ray spectrometers on multiple OSOs, to my knowledge, no data was ever exchanged between Ken Frost and Larry Peterson. Except for their one collaborative balloon flight of the first actively shielded spectrometer \citep{1962ITNS....9..381F}, they went their separate ways, Ken at Goddard and Larry at the University of California San Diego, and they never collaborated again. There were similar stand-offs in solar radio astronomy with Mukul Kundu's group at the University of Maryland and Hal Zirin's group at Cal Tech.

Perhaps the most contentious field was in solar gamma-ray observations and their interpretation.  Ed Chupp was the PI of the team at the University of New Hampshire that had made the first pioneering observations of solar gamma-rays \citep{1973Natur.241..333C}.  Ed was perhaps the strongest believer in PI data rights and he strictly controlled who had access to data from his instruments. Even Reuven Ramaty had difficulty getting access to the data for comparisons with the predictions of his extensive modeling work.  Reuven with Richard Lingenfelter had developed the theory of gamma-ray production in solar flares \citep{1975SSRv...18..341R}. They amassed an extensive data base of nuclear cross-sections for the production of gamma-rays from different interactions of accelerated ions with ambient nuclei of all different elements. Along with his postdoc, Natalia Mandzhavidze, and his graduates students, particularly Ron Murphy, Reuven developed a comprehensive Fortran program to use this data base to calculate the expected gamma-ray spectrum from solar flares for various assumed spectra and compositions of the accelerated particles. But Reuven was also strongly opposed to open access to this code and to the results that were obtained from it. This, of course, led to a delicate relationship with Ed Chupp.  Reuven could only compare his predictions with the gamma-ray spectra that Ed had published and Ed could only use the predictions that Reuven had published to compare with his measured spectra. 

This situation continued into the 1980s and '90s. The situation got so bad that even Ed's colleague, Dave Forrest at the University of New Hampshire, was driven to take Ed to court even though he was a critical part of the team that designed and built the gamma-ray spectrometer in the first place.

Ken Frost was originally also opposed to open access to the data from his instruments on the OSOs and on SMM.   He felt, like Carl Fichtel, that it was better to keep everything within his group at Goddard including being fully responsible for all aspects of the instrument itself and maintaining full control of all access to the data. In this way, he could be sure that the instrument would be optimally designed, fabricated, calibrated, and operated according to his specifications, and he could establish his reputation based on the published results. He felt that he had to control access to the data from the instrument so that he would be sure to get recognition and credit for the results and any new discoveries. Establishing and maintaining his reputation was clearly important in continuing to obtain funding for future projects although, in those days, the competition was much less intense than it is today. Also, decisions on the selection and funding of new instruments was made as more like ``gentleman's agreements'' rather than by the peer review process we have today. This was how there came to be X-ray spectrometers on alternating OSOs by the Frost and Peterson teams. But a more altruistic argument could be made for controlled data access in that it could prevent false or inaccurate results from being published by fellow scientists who may be unqualified or unfamiliar with the instrument and the vagaries of the observations.

The situation began to change with the launch of SMM in 1980.  There were still strong PIs for each of the different instruments but it was being recognized that scientific progress depended more on combining observations in different regimes - between thermal and nonthermal emissions, for example, or between hard X-rays from accelerated electrons and gamma-rays from protons and heavier ions.  Furthermore, most of the PIs and many of their team members spent a considerable time co-located at Goddard to control the operation of their instrument and to have easy access to the data. This greatly facilitated collaborations between the different teams, initially through the joint planning of coordinated observations following the daily planning meetings in which each team got one vote in deciding on the observing plan for the next day. Comparing observations from the different instruments also became commonplace, and the presence of many scientists from around the world in Goddard's Building Seven during the first year of operations provided me with a wonderful opportunity to learn, what was for me at the time, a new field of research in solar physics. We were all trying to make progress in a new open and collaborative atmosphere as compared to the old closed and competitive approach.

I am pleased that I was able to persuade Ken Frost to make the data from his instrument on SMM, the \textit{Hard X-ray Burst Spectrometer} (HXRBS), completely open for access and analysis. Perhaps this was partly motivated by the fact that it was a relatively simple data set with the count rate in 15 energy channels from $\sim$20 to 500 keV read out every 128 ms. Nevertheless, the free availability of the HXRBS data meant that most of the other SMM instrument teams used HXRBS light curves for individual flares as a standard, much like the GOES soft X-ray light curves are used today.  The HXRBS light curves at different energies are featured in the analysis of the major solar flare of 21 May 1980 that \cite{1985SoPh..100..435D} said in their review paper ``was one of the best observed, and best analyzed flares in the whole history of solar research. Its observations on ...SMM.. led to the first study of images of hard X-ray brightenings at footpoints of loops during the impulsive phase \citep{1981ApJ...246L.155H} 
and to the discovery of giant post-flare coronal arches following two-ribbon flares \citep{1982SoPh...75..305S}.'' The HXRBS data set, ultimately including over 12 000 flares \citep{1991chxb.book.....D}, became far more useful in understanding the importance of flare accelerated electrons than it would have if it had been restricted to analysis by members of the PI team alone.  Also, I think the widespread use of HXRBS data cemented Ken Frost's reputation and led in no small measure to the continued success of the solar X-ray group at Goddard. But perhaps of equal importance, it was a pioneer in leading the way to the open data access that we enjoy in solar physics today.

The fears of open data access by the early PIs have proven to be groundless. There is much greater competition now for building new instruments than ever before, despite the knowledge that the instrument builders will not have privileged data rights.  While a few false or inaccurate results have been published by scientists unfamiliar with the detailed instrument characteristics, they are overwhelmed by the increased number of excellent papers resulting from the larger numbers of people who now have free access to the data.  The joint analysis of observations by widely different instruments has vastly increased, as has comparisons between observations and predictions of various theoretical models and simulations.  I believe that this has led to a more efficient use of the available data and a more rapid advance in our scientific understanding than would otherwise have occurred.

\section{\textit{Orbiting Solar Observatories}}
    \label{S-OSOs}

Many solar satellites have been flown since the dawn of the space age, and some continue to fly today  and the instruments continue to operate, some after 20 or more years (\textit{e.g.} SOHO).\footnote{http://www.scholarpedia.org/article/Solar\_Satellites}

Early in the space program, under the direction of John Lindsay, the head of the Solar Physics Branch at Goddard Space Flight Center, NASA initiated the Orbiting Solar Observatory (OSO) series of dedicated unmanned solar satellites. Together with Skylab, the solar instruments carried on these early spacecraft provided outstanding UV and EUV imaging and spectroscopic observations together with X-ray, and gamma-ray spectroscopy. The results of these pioneering observations still, to this day, provide the foundation for much of solar space science.

The OSO series eventually involved a total of nine spacecraft. They were given sequential letter designations (OSO-A through OSO-I) until launch at which point they were renamed with the next number in the sequence. OSO-C was not launched because of a failure during testing on 14 April 1964 that damaged the spacecraft and resulted in the death of three engineers. Thus, the eight successful spacecraft were named OSO-1, launched in 1962, through OSO-8, launched in 1975. The first seven OSOs were built by Ball Brothers Research Corporation (later called Ball Aerospace) but Hughes Aircraft won the follow-on contract for OSO-I, J, and K. (Ken was known to favor the Ball Brothers proposal and that's how he became known as \it ``a ball supporter.'') \rm OSO-I became OSO-8 at launch on 21 June 1975 but \hbox{OSO-J} and -K were combined together for budgetary reasons to become SMM (see Section \ref{S-SMM}). 

\begin{figure}
	\begin{center}
    \includegraphics*[width=0.7\textwidth, angle = 0, 
   trim = 0 0 0 0]
   		{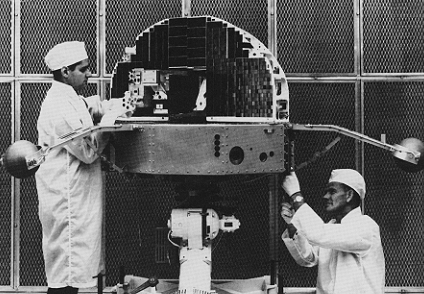}
    \caption{One of the early OSOs being prepared for testing showing the lower wheel section that spun at up to 15 rpm with the de-spun section above it. Two of the three arms are shown with each with a ball filled with propellant for attitude and spin control.}
   \label{Fig-OSO}
   \end{center}
\end{figure}

As shown in Figure \ref{Fig-OSO}, all the OSO spacecraft incorporated a ``wheel'' component spinning at up to 15 rpm to provide pointing stability, and a de-spun platform to carry the imaging instruments pointed at the Sun. This arrangement allowed for sub-arcsecond pointing accuracy and stability before the era of 3-axis stabilization techniques that are used today. Several other instruments that did not need to be constantly and precisely pointed at the Sun were mounted in the spinning ``wheel'' section. These included X-ray and gamma-ray spectrometers pointed perpendicular to the spin axis so that they scanned across the Sun each rotation, and similar instruments pointed parallel or near-parallel to the spin axis to view non-solar sources of astrophysical interest. Solar panels mounted on the de-spun platform provided the power, and slip-rings between the spinning and de-spun components were used for power and signal connections.

\begin{figure}
  	\begin{center}
    \includegraphics*[width=1.0\textwidth, angle = 0, 
   trim = 0 0 0 0]
   		{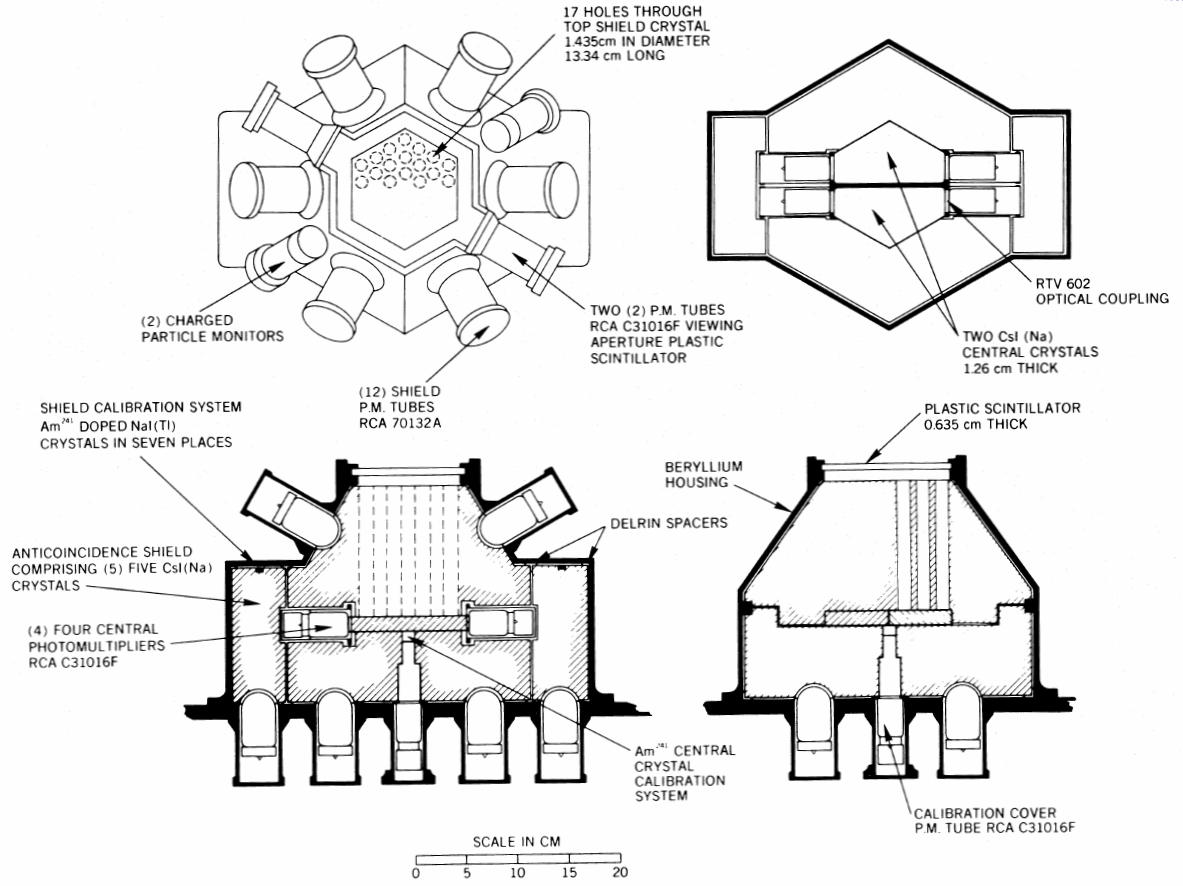}
    \includegraphics*[width=0.5\textwidth, angle = -90, 
   trim = 120 0 50 0]
   		{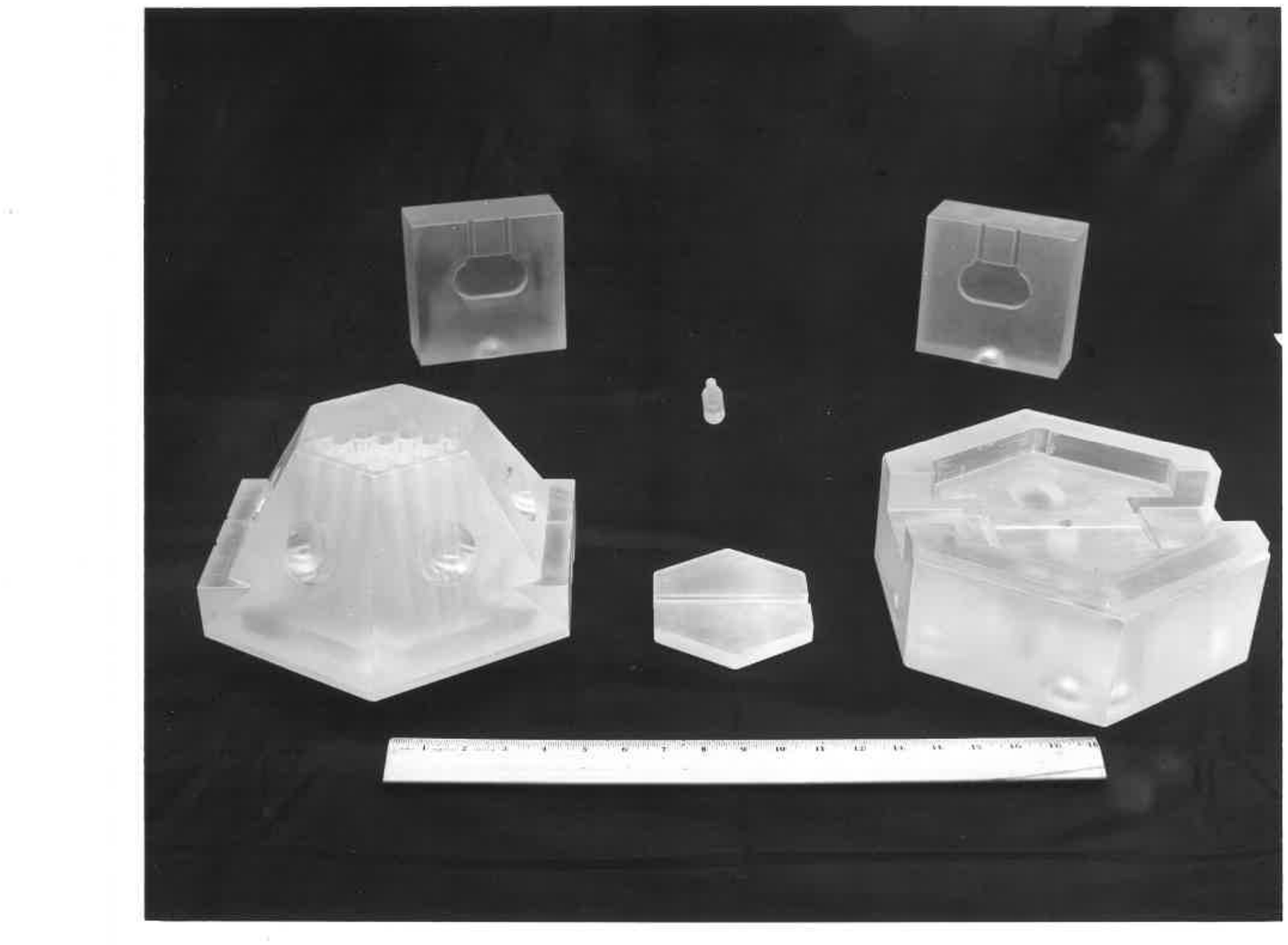}  
    \caption{\textit{Top}: Diagram of the OSO-8 \textit{Celestial X-ray Spectrometer}. \textit{Bottom}: Photograph of the bare CsI(Na) crystal components.       }
   \label{Fig-OSO8}
   \end{center}
\end{figure}

The \textit{Celestial X-ray Spectrometer} on OSO-8 shown in Figure \ref{Fig-OSO8} \citep{1975...X68275159F, 1977SSI.....3..325D, 1977ApJ...217..809D} represented our best effort at getting a detector with the lowest background possible using active anticoincidence shielding. The primary objectives were to measure the energy spectrum of cosmic X-ray sources above 20 keV and to search for time variations in their intensity. The detector consisted of two optically isolated central crystals shielded by 2-inch thick crystals and a large, active collimator. The sensitive area was 27.5 cm$^2$ and the field-of-view was $5^{\circ}$ FWHM. The instrument was mounted in the wheel section of OSO-8 with the axis of its field of view offset by $5^{\circ}$ from the negative spin axis of the wheel. The minimum detectable intensity of a point source which is brought to within $5^{\circ}$ of the negative spin axis for greater than one day was $\sim10^{-5}$ photons~cm$^{-2}$ s$^{-1}$ at 100 keV.  The bare CsI(Na) crystals that made up the sensitive volume of this instrument are shown in Figure \ref{Fig-OSO8}.

\section{The Solar Maximum Mission (SMM)}
    \label{S-SMM}

\begin{figure}    
	\begin{center}
    \includegraphics*[width=0.7\textwidth, angle = 0, 
   trim = 0 0 0 0]
   		{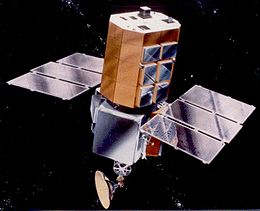}
    \caption{\textit{The Solar Maximum Mission} (SMM), launched in 1980, repaired in orbit in 1984, and re-entered in 1989.}
   \label{Fig-SMM}
   \end{center}
\end{figure}

In contrast with all of the OSO spacecraft, SMM, shown in Figure \ref{Fig-SMM}, was three-axis stabilized so that all of the instruments were pointed at the Sun with arc-second accuracy and stability. The primary scientific objective of the mission was to further our understanding of solar flares. These powerful phenomena had been revealed by earlier observations to be the most energetic explosions in the Solar System but the physical processes involved in the impulsive energy release and subsequent dissipation were largely unknown. To this end, imaging and spectroscopy was carried out at UV, EUV, and X-ray wavelengths to study the emission from plasma heated to temperatures as high as several tens of million kelvin. The electrons accelerated during flares to suprathermal energies (basically $>$10~keV) were detected through the bremsstrahlung X-rays that they produced. SMM included the first instrument capable of imaging these X-rays up to energies of 30 keV, well above the energies produced by all but the hottest plasma. The protons and heavier ions also accelerated during flares were detected through the nuclear gamma-rays that they generate as they interact in the solar atmosphere. SMM carried a sensitive gamma-ray spectrometer similar to the spectrometer that had made the first pioneering measurements of this high energy emission on OSO-7 back in 1972 \citep{1973Natur.241..333C}. Reviews of the scientific results from SMM can be found in \cite{1999mfs..conf.....S}.

SMM was launched by a Delta rocket on 14 February 1980, on the rising phase of the 11-year cycle of activity. It operated successfully until November 1980, when the last of several fuses in the aspect control system failed and the spacecraft was no longer able to maintain the orientation towards the Sun with the arc-second stability that was needed by the imaging instruments. The non-imaging instruments, HXRBS included, continued to operate as the spacecraft slowly spun about an axis pointed within a few degrees of Sun center. Then in April 1984, in the first of many dramatic spacecraft rescue and servicing missions using the Space Shuttle, the aspect system was fixed by the astronauts, and SMM was able to provide six more years of excellent solar observations until it re-entered the Earth's atmosphere in December 1989.

\section{The Road to RHESSI}

The story of how the concept of solar hard X-ray and gamma-ray imaging spectroscopy that started in the 1970s led to RHESSI over thirty years later is summarized in the 100th RHESSI Nugget.\footnote{http://sprg.ssl.berkeley.edu/$\sim$tohban/wiki/index.php/Nugget\_100} It was a long hard road and consumed much of my time and that of many other people, notably, Bob Lin and the scientists who were to become CoIs on the final winning SMEX mission. But that it was ultimately so successful is a testimony to the quality of the work that went into it and an excellent example of how persistence can be rewarded under the right circumstances.

Hard X-ray imaging started with the \textit{Hard X-ray Imaging Spectrometer}\linebreak[4] (HXIS) \citep{1980SoPh...65...39V} on SMM, launched in 1980. It used an instrument with multiple pin-hole collimators and was capable of imaging in X-rays up to 30 keV with 8 arc-second pixels. \textit{Hinotori}\footnote{http://www.isas.jaxa.jp/en/missions/spacecraft/past/hinotori.html}, launched in 1981, followed soon thereafter using rotating modulation collimators \citep{1978nisa.symp..277M,1982sofl.symp..120M} to achieve similar angular resolution but greater sensitive area. And then of course, the \textit{Hard X-Ray Telescope} (HXT) \citep{1991SoPh..136...17K} on \textit{Yohkoh} in the 1990's improved on the angular resolution and extended the energy range up to almost 100 keV.

The  \textit{Pinhole Occulter Facility} (P/OF - pronounced ``poff'') was proposed in the 1980s \citep{1989SPIE.1159..318H}. It would have used a boom that could be extended to 50 m to achieve sub-arcsecond resolution using bi-grid collimators. P/OF was initially proposed for the \textit{Space Shuttle} and later for the \textit{Space Station} but was never selected for funding. The less ambitious balloon-borne \textit{High Energy Imaging Device} (HEIDI), initiated by Carol Crannell\footnote{https://en.wikipedia.org/wiki/Carol\_Jo\_Crannell} at Goddard, had two rotating modulation collimators and a solar aspect system concept similar to that used on RHESSI. It was funded but was flown only once from Texas in 1993 \citep{1991ibt..conf....1C}. (The original HEIDI instrument paper was presented at the AIAA international Balloon Technology Conference held in 1991 in Albuquerque, NM, but unfortunately, it is not available on line.  The results from the balloon flight were presented at the AIAA $32^{nd}$ Aerospace Sciences Meeting and Exhibit held in 1994 in Reno, NV. It is available on line\footnote{https://arc.aiaa.org/doi/10.2514/6.1994-299} but it is not listed in NASA ADS.)

\subsection{\textit{Solar High-Energy Astrophysical Plasmas Explorer} (SHAPE)}

Loren Acton suggested that P/OF should be put on SMM during a second repair mission in the late 1980s.  This was never taken too seriously and was probably technically infeasible but it did lead to SHAPE, which included greatly upgraded versions of the SMM flare capabilities and the high-angular-resolution hard X-ray imaging envisioned for P/OF.

\begin{figure}
	\includegraphics[width = 1.0\textwidth]{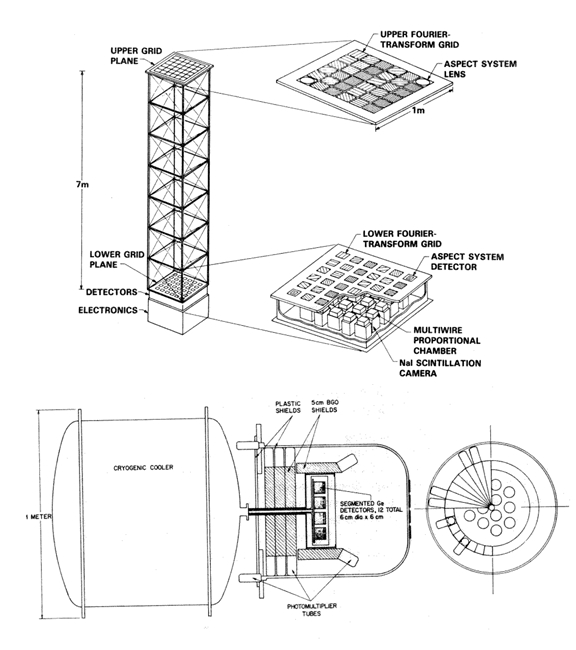}
    \caption{The two main instruments included in the proposed \textit{Solar High-energy Astrophysical Plasmas Explorer} (SHAPE). The \textit{Gamma-Ray Imaging Device} is shown on the top and the \textit{High-resolution Gamma-Ray, Hard X-ray, and Neutron Spectrometer} (HiGRANS) on the bottom.}
    \label{Fig-SHAPE}
\end{figure}

The SHAPE proposal, submitted to NASA in July 1986, was to fly four instruments to study the high-energy aspects of solar flares \citep{1986STIN...8725983D}.  Amazingly, it is still available online\footnote{https://ntrs.nasa.gov/search.jsp?R=19870016550}. It grew out of the Explorers Concept Study Program started in 1986 and was considered as the logical follow-on to the SMM flare observations. How I became the PI of this abortive attempt to fly the best X-ray, gamma-ray, and UV instruments of the time, all on the same spacecraft, is an interesting tale of how to cope with a group of brilliant and ambitious scientists, all wanting to be the PI for their own instrument.  I wasn't really promoting any particular instrument other than my involvement in the field as the PI for the HXRBS instrument on SMM since Ken Frost stepped down from that role soon after launch. So by popular agreement, I was chosen to lead this disparate group.

The scientific objectives read much the same then as they do today for more recent proposals - understanding energy release and particle acceleration, flare geometry and energy transport, Solar abundances, non-flaring studies, etc. The main contenders for the instruments to be proposed were the \textit{Gamma-Ray Imaging Device} (GRID), a Fourier-transform X-ray and gamma-ray imager with arc-second angular resolution similar to the imager proposed for P/OF \citep{1989SPIE.1159..318H} and led by Tom Prince from CalTech; the \textit{High-resolution Gamma-Ray and Neutron Spectrometer} (HIGRANS) \citep{1988AIPC..170..456L}, a cooled germanium spectrometer with keV energy resolution led by Bob Lin at the University of California, Berkeley; the \textit{Soft X-ray Impulsive Phase Spectrometer} (SIPS), a Bragg-crystal X-ray spectrometer led by George Doschek at NRL; an X-ray polarimeter led by Bob Novick from Cornell; and the \textit{Low-Energy Imaging Spectrometer} (LEIS), an imaging spectrometer with arc-second spatial resolution operating at UV and EUV wavelengths, and led by Marylin Bruner from Lockheed. Drawings of GRID and HIGRANS are shown in Figure \ref{Fig-SHAPE}. After much discussion, the X-ray polarimeter was dropped as being too risky with limited and uncertain scientific return.

Sadly, this beautiful mission was not to be. It was caught up in the hiatus at NASA following the Challenger accident on 28 January 1986.  According to Len Fisk, the Associate Administrator for Space Science and Applications at the time, in a letter dated 4 September 1987, ``Its effects and other program adjustments have resulted in a delay in development of new missions in the Explorer Program by approximately 2 to 3 years.''  By the time it was reviewed, it was deemed to be too late to be in time for the upcoming maximum of solar activity, and thus, despite getting excellent ratings in all other categories, it was not selected, and we were faced with an 11-year delay before another mission could be realistically proposed.

In writing this memoir, I discovered a paper that I had forgotten written by Reuven Ramaty, Gordon Emslie, and myself \citep{1988SoPh..118...17R}.  Since Reuven was the lead author, it emphasizes the case for gamma-ray and neutron observations but apparently Gordon and I got in about equal time for hard X-ray studies.  We listed the basic requirements for a high-energy solar physics facility that matched pretty well what had been the SHAPE proposal. This is the sort of paper you have to write when you are between missions and doing everything you can to keep future plans in play. There have been only 11 citations to this paper so it is not clear if it had any significant impact at the time but it does make for interesting reading even now.

\subsection{The Max91 Program}

I started what became known as the Max91 Program in frustration after the SHAPE proposal was rejected. I realized that we were not going to have a solar-dedicated space mission for the next solar cycle and that the best hope was to marshal the opportunities that might be available with astrophysics space missions, possible balloon-borne solar instruments, and existing ground-based observatories. The first thing I did was to call Dick Canfield to see if he was interested in helping to organize it. In his usual pragmatic way, he focused on the most important objective of any effort like this, which is to obtain more funding for solar physics from NASA and NSF.  My more altruistic and perhaps naive idea was that it would be good for the science if future observations were coordinated so that the same flares could be observed across the electromagnetic spectrum. In any case, we were quite successful in getting support from the funding agencies and also from the solar physics community. We quickly got help from Ernie Hildner at NOAA, and eventually Alan Kiplinger took on the role of running the program itself for several years. We wrote several mission papers, organized several workshops, and I think helped get funding for the two balloon programs that were successful in flying during the peak of Cycle 22 - Bob Lin's HIGRANS X-ray/gamma-ray spectrometer \citep{1988AIPC..170..456L,1988fnsm.work..262L} and Carol Crannell's HEIDI X-ray imaging spectrometer. 

With the selection of HESSI in 1997, Dick Canfield transitioned the then somewhat moribund Max91 program into the Max Millennium program\footnote{http://solar.physics.montana.edu/max\_millennium/}
that has been invaluable in supporting the RHESSI mission and is still operating today. It is currently moderated by Dick and his colleague, Keiji Yoshimura, at Montana State University. A group of volunteers (Shaun Bloomfield, Northumbria University; Ying Li, Nanjing University and  Montana State University; William Marquette, Helio Research; Ryan Milligan, Queen's University Belfast) act as the Max Millennium Chief Observer and keep abreast of solar activity on a daily basis. They  select the most appropriate observing targets and observing plans using data from observatories around the world and their expertise at solar activity forecasting, and send out their recommendations to hundreds of solar scientists via the Max Millennium Message of the Day (MMMoD).  As of this writing, 6491 messages have been sent since it started on 11 June 1999. 

\subsection{Astrophysics Missions}

Although strictly speaking not a solar satellite, the \textit{Compton Gamma Ray Observatory} (CGRO), the second of NASA's \textit{Great Observatories}, made many important observations of solar flare X-rays and gamma-rays. Launched in April 1991 and re-entering the Earth's atmosphere in 4 June 2000, CGRO covered almost a solar cycle with observations spanning an unprecedented six decades of the electromagnetic spectrum, from X-rays as low as 30 keV to gamma rays at 30 GeV. 



I had a CGRO Guest Investigator project funded for several years early in the mission, and we made the solar flare data from the \textit{Burst And Transient Source Experiment} (BATSE) (spanning energies from $<$10 keV to $>$10 MeV) readily available on line at the Solar Data Analysis Center.\footnote{https://umbra.nascom.nasa.gov/index.html/} Richard Schwartz and Kim Tolbert made sure that the OSPEX software package developed to analyze RHESSI spectroscopy observations could also handle BATSE data. We did a similar thing with the data from the \textit{Fermi Gamma-Ray Burst Monitor} (GBM). I think that the relatively easy access to the solar data from these astrophysics missions and the use of familiar analysis tools greatly enhanced the scientific usefulness of these important data sets and allowed more people to include them in their solar research. 

More recent gamma-ray observations continue to be made with the even more sensitive ESA \textit{INTErnational Gamma-Ray Astrophysics Laboratory} (INTEGRAL) in cooperation with Russia and the United States. The four instruments on INTEGRAL cover the energy range from 3 keV to 10 MeV. Unfortunately, INTEGRAL cannot be pointed at the Sun, and so flare gamma-rays can only be detected through the thick anticoincidence shield around the high-resolution germanium detectors.

\subsection{The \textit{High Energy Solar Physics} (HESP) Study}

During the early 1990s, some of the original SHAPE team, led by Bob Lin, began holding meetings to plan for the future of high energy solar physics. This became known as the HESP study, a name that we stole from the Japanese, who had used it as the original name for \textit{Yohkoh}.  The study team quickly focused on the idea of combining the SHAPE X-ray imager and the high resolution spectrometer into a single instrument. I'm not sure who first came up with this idea but I suspect Bob was heavily involved, if not the originator himself. Combining the functions of the two instruments would be achieved by using tungsten grid pairs with different slit pitches to form modulation collimators in front of multiple cooled intrinsic germanium detectors.  It was also realized that improved imaging would be obtained by rotating the instrument about the collimator axis to measure many more spatial Fourier components than could be achieved with a pointed instrument like the \textit{Hard X-ray Telescope} (HXT) imager on \textit{Yohkoh} \citep{1991SoPh..136...17K}. The easiest way to rotate the instrument was to spin the whole spacecraft but this precluded the possibility of including the other conventional UV imagers of the original SHAPE proposal since they required 3-axis pointing.  It was felt that the advantages for X-ray and gamma-ray imaging of spinning the spacecraft outweighed the loss of the conventional imagers on the same spacecraft. They could be flown on their own separate missions and, along with ground-based observatories, would still provide simultaneous context observations of the thermal emissions that would be critical for interpreting the X-ray and gamma-ray data.

\begin{figure}
	\includegraphics[width = 1.0\textwidth]{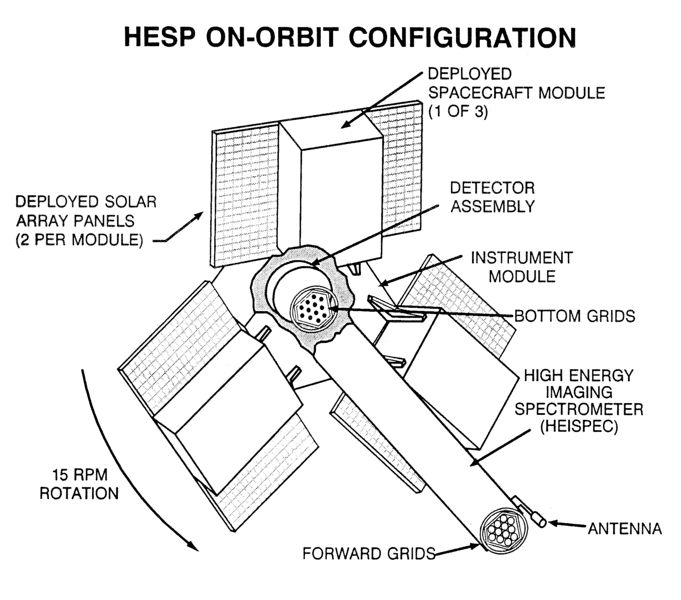}
    \caption{The High Energy Solar Physics (HESP) mission concept.}
    \label{Fig-HESP}
\end{figure}

This concept of a single instrument capable of both X-ray and gamma-ray imaging with high angular resolution and spectroscopy with high energy resolution was further developed in the early 1990s \citep{1990SPIE.1344..492H, 1996SPIE.2804..228D, 1998SPIE.3442....2L}. The spectroscopy was achieved with hyper-pure cooled germanium detectors (HPGes) developed under the direction of Dick Pehl at the Lawrence Berkeley Laboratory. The imaging was achieved using multiple bigrid collimators, one in front of each of the germanium detectors, with the whole spacecraft rotating at $\sim$15~rpm to produce the rapid modulation of the incident solar X-ray and gamma-ray fluxes that provided measures of many spatial Fourier components of the sources on the Sun. The combined instrument was capable of high-resolution spectroscopy at each location in the image for a flare that occurred anywhere within $\sim$$1^{\circ}$ of the spacecraft spin axis that was generally to be pointed within a few arcminutes of Sun-center. This concept is best explained in the instrument papers that were published after HESSI was finally launched in 2002 \citep{2002SoPh..210....3L,2002SoPh..210...33S,2002SoPh..210...61H,2002SoPh..210...87F}.   Various versions of this concept were proposed to NASA for different opportunities.  These included the HESP mission shown in Figure \ref{Fig-HESP}. It was proposed in 1991 as a medium-class satellite by a science study group of the same name  chaired by Bob Lin. It was not selected for funding but it was the first time that a combined imager/spectrometer was proposed as a single instrument with the same basic elements as in the final HESSI SMEX proposal, a major step towards the first imaging spectroscopy in the hard X-ray/gamma-ray spectral domain.

\subsection{The HESI Study and the HESSI MIDEX Proposal}
	\label{s-HESI-study}

HESP was quickly followed by the \textit{High Energy Solar Imager} study (HESI) and by the resulting \textit{Medium-class Explorer} (MIDEX) proposal in 1995 given the name HESSI with the second ``S'' added to acknowledge the importance of spectroscopy. The basic concept was initially put forward for a major Solar mission opportunity with a nearly half billion dollar price tag. The proposed instrument had 36 bi-grid modulation collimators each feeding a 7-cm diameter germanium detector, the largest available at the time. It was beaten out by the \textit{Thermosphere Ionosphere Mesosphere Energetics and Dynamics} (TIMED) mission because, as we later learned, George Withbroe, the NASA selecting official at the time, wanted to have a full set of spacecraft covering all locations from the Sun to the Earth. Undaunted, we proposed a new descoped version with just 12 collimator/detectors for the MIDEX opportunity a year later but again it was beaten out by TIMED since the major Solar mission line had not been funded.  Withbroe said at one of the many presentations that I made at NASA HQ that our mission should be renamed \emph{``After TIMED.''} It was during this depressing period of repeated rejections and  regular presentations of various versions of our proposal that I developed my deep dislike of interacting with NASA HQ, and my firm commitment to never work there if I could help it.  Fortunately, apart from serving on review panels for various research proposals, I have succeeded in keeping that commitment.

\subsection{HESSI SMEX}
	\label{S-HESSI_SMEX}

Our perseverance was finally rewarded in 1997, when our further descoped mission with just nine RMC/detectors was accepted as a SMEX. We managed to fit within the SMEX budget, about half the MIDEX cap, by enlisting support from the Paul Schirrer Institut (PSI) in Switzerland. By a stroke of good fortune, Bob Lin had spent time visiting Arnold Benz in Z\"{u}rich the previous year and had met with Alex Zehnder at PSI. It was agreed that they would obtain funding from the Swiss government to pay for most of the collimator integration and alignment work.  This reduced the cost to NASA by about \$2 million in our proposed budget although it would probably have cost more like \$10 million if the work had been done at Goddard as we had planned to do for the MIDEX version. This collaboration turned out to be very successful and mutually rewarding in many ways.  For one thing, it meant that several of us spent two three-week periods in Switzerland helping to mount and align the grids and doing optical and X-ray tests of the alignment. It also led to close collaborations with Arnold Benz and his  excellent students, many of whom are still working on RHESSI both in Switzerland and in the US. These include the current RHESSI PI, S\"{a}m Krucker, Markus Aschwanden, Martin Fivian, Pascal Saint-Hilaire, Marina Battaglia, and others

Further \textit{difficulties} were encountered, even after the SMEX proposal was accepted. By early 2000, we were within budget and well on our way to meeting the planned launch date of 4th July, 2000, when the now infamous disaster struck during the final full-up spacecraft vibration test at the Jet Propulsion Laboratory (JPL). According to the investigation committee's finding, the vibration table had broken during its previous use several months earlier. Apparently, it was jammed when the HESSI spacecraft was first mounted on it.  When the command was sent to start the vibration test, the automatic control system was set to ramp up the input until it sensed the planned vibration level.  The input was increased until the table suddenly freed itself and hit the spacecraft with a vibration level of $\sim$20 G, several times higher than planned.  This broke the solar panels and two of the three kinematic mounts holding the telescope tube in place. If the third mount had broken, the whole telescope would probably have been destroyed. Fortunately, Ed Weiler, the responsible NASA official, did not see the photographs showing the full extent of the damage and agreed to provide an additional \$10 million to repair the instrument. Much of this had to be given to the Swiss since the laborious process of mounting and aligning the grids had to be repeated. It took us about a year to get ready again for launch.  

Unfortunately, problems with the Pegasus rocket delayed the launch yet again for a further nine months or so. Even on the day of the launch itself, 4 February 2002, a further delay was caused by an open microphone in the L1011 plane carrying the Pegasus over the Atlantic to the drop site. The plane had to circle back around the ``racetrack'' path before dropping the Pegasus half an hour later.

Finally, after a perfect launch and insertion into a near-circular orbit at the desired altitude, HESSI has never looked back. In a perverse way, it may have benefited from all of the descopes, delays, mishaps, and frustrations.  As Ken Frost would have said, ``They were all maturing experiences.'' I know I felt much more \emph{``mature''} and relieved once it was in orbit.

\section{The RHESSI Era}
	\label{S-RHESSI_Era}

This section should have been the easiest for me to write because in a very real sense, my work with RHESSI has been the culmination of my career. I have been able to use all of my innate capabilities along with the skills learned through all of my formal academic education and the extensive on-the-job training on several earlier missions. The time since RHESSI's launch has been unquestionably the most satisfying of my career.  While perhaps not as educational for me as the SMM operations decade of the 1980s, when I was younger and new to solar physics, RHESSI has certainly been the most productive mission that I have been involved with.

Even with its delayed launch past solar maximum, it quickly achieved its minimum mission requirements as required by NASA to be declared a success.  By early 2003, after only a year in orbit, we reported in the first of the RHESSI Senior Review proposals, which NASA requires for all operating missions every two years, that over 8,000 solar X-ray events had been recoded. The Senior Review Panel stated in their report that \emph{``Early observations with RHESSI have revealed information on flare energetics, timing and spatial structure that will simulate renewed efforts to model and understand flares and magnetic reconnection on the sun.''} Even as solar activity died down leading to solar minimum in 2008, the panel agreed that \emph{`RHESSI has a high potential for future ``discovery'' observations.'}

The biennial Senior Review proposals provide a convenient summary of the RHESSI scientific productivity throughout its now 16-year lifetime\footnote{https://hesperia.gsfc.nasa.gov/rhessi3/mission/documents/index.html}.  While I regularly complained that these proposals were too frequent and required a significant amount of my time and the time of other RHESSI team members to prepare, I now recognize that they have historical value if nothing else. In addition to the 20+ page proposal and budget, a half-hour presentation had to be made to the Senior Review panel each time. Bob Lin would give this presentation in his usual somewhat frenetic pace with close to 50 PowerPoint slides to impress the panel with the number and breadth of the scientific results. We usually got pretty good reviews as a result.  Only in April 2010 did I have to give the presentation when Bob was trapped in London following the eruption of the Icelandic volcano, Eyjafjallajökull, which led to air travel disruptions because of volcanic ash in the atmosphere. Since Bob's death in 2012, S\"{a}m Krucker, the new RHESSI PI, has given the presentation each year with similar success in getting funding for continuing operations.

Early RHESSI scientific results were also published along with the instrument papers in the special issue  of \textit{Solar Physics} (Volume 210, 2002). In retrospect, this series of 26 papers, published in less than a year after launch was a remarkable achievement.  The instrument papers are still cited today with over 1000 citations for \cite{2002SoPh..210....3L}, while the paper I was most heavily involved with \citep{2002SoPh..210..341G} has 97 citations. The latter paper covered the first GOES X-class flare observed with RHESSI, the X1.5 flare on 21 April 2002. I became somewhat obsessed with this flare since it was so well observed with RHESSI and was a classic two-stage event with a nonthermal impulsive phase and a thermal long-duration phase visible in RHESSI soft X-ray images between 3 and 12 keV for over 12 hours as the coronal source moved away from the Sun. Many years later, I published a paper on a similar long-duration event that showed intriguing quasi-periodic pulsations (QPP) for several hours after the impulsive phase \citep{2017ApJ...836...84D}.

Another group of 14 RHESSI papers on the intense (X4.8) gamma-ray flare of 23 july 2002 was published in the special issue of \textit{Astrophysical Journal Letters} (Vol. 595, 2003). The RHESSI hard X-ray and gamma-ray observations are summarized in \cite{2003ApJ...595L..69L}.  This was the first flare that allowed all of RHESSI's capabilities to be used.  It provided the first gamma-ray images of a solar flare \citep{2003ApJ...595L..77H} and the first solar flare measurements at high energy resolution of nuclear de-excitations lines \citep{2003ApJ...595L..81S} and the positron-electron annihilation line \citep{2003ApJ...595L..85S}.

At the time of RHESSI's 10$^{th}$ anniversary in 2012, Bob Lin and I summarized the major scientific results in a Science Nugget.\footnote{http://sprg.ssl.berkeley.edu/$\sim$tohban/wiki/index.php/RHESSI's\_Tenth\_Anniversary}  As a result of a pseudo-democratic process, the following top-ten list of flare related items, in priority order, was approved by a majority of many members of the RHESSI PI team:

\begin{enumerate}
  \item Discovery of Gamma-Ray Footpoint Structures
  \item Energy Content and Spectrum of Flare Energetic Electrons
  \item Ubiquitous Nonthermal Emissions from the Corona, and Bulk Energization
  \item Double Coronal X-ray Sources
  \item Microflares and the Quiet Sun
  \item Initial Downward Motion of X-ray Sources
  \item HXR Flare Ribbons
  \item Location of Superhot X-ray Source
  \item Photosphere as a Compton or "Dentist's" Mirror
  \item Broadened 511-keV Positron Annihilation Line
\end{enumerate}

A further list of three non-flare-related areas were identified that RHESSI had contributed to:
\begin{enumerate}
  \item Solar Oblateness
  \item Magnetar Timing and Spectroscopy
  \item Terrestrial Gamma-ray Flashes (TGFs)
\end{enumerate}

\subsection{The Secret to RHESSI's Success}

What was the secret to the extraordinary scientific success of RHESSI? That is the question I will try to answer in this section.  Perhaps my attempt will be helpful to people working on future missions.

The answer to RHESSI's success starts with the PI, Bob Lin.  As I said at the emotional meeting in Sonoma following Bob's tragic death in 2012, without him, quite simply, RHESSI would not have happened.  To make the point as strongly as I could, I paraphrased the comment made by Texas Senator Lloyd Bentsen in a 1988 vice-presidents debate to the Republican candidate Senator Dan Quayle in response to Quayle comparing himself to John F. Kennedy.  I said that if anybody claims that he can take over from Bob, I will tell them that "I knew Bob Lin. He was a friend of mine and you are no Bob Lin." I also said that the only way to continue on without Bob was for all of us to work together, each person making his or her own unique contributions, and hope that the sum of our efforts will be half as good of what Bob would have achieved if he had lived on.

The second critical factor leading to RHESSI's success was the fortuitous coming together of several new techniques and new technology that gave the mission several truly ``first ever'' capabilities. In the annoying words of the NASA Headquarters people, these were revolutionary, not just evolutionary advances.  Thus, in the 1997 HESSI SMEX proposal, we were able to make the following statement: 

HESSI will provide 
\begin{itemize}
  \item{the first hard X-ray imaging spectroscopy,}
  \item{the first high-resolution spectroscopy of solar gamma-ray lines,}
  \item{the first imaging above 100 keV, and}
  \item{the first imaging of solar gamma-ray lines.}
\end{itemize}
All of these firsts were subsequently achieved in the first couple of years after launch.

These new capabilities were the result of many years of development both in the lab.~and on many balloon flights.  They included advances in both imaging techniques and spectroscopy, either one of which, by itself, would have been worthy of a SMEX but together they were an unbeatable combination.  They allowed true imaging spectroscopy with a high-resolution spectrum at each pixel of the image. In layman's terms, we could make color X-ray movies of solar flares.  I tried to claim that we would have a similar dramatic effect on solar physics as the movie, \emph{The Wizard of Oz}, had when it switched from black-and-white to color but that argument didn't seem to impress anybody. In addition to the X-ray imaging spectroscopy giving us information about the accelerated electrons and the hottest thermal plasma, we also had gamma-ray imaging spectroscopy with information about the accelerated ions. Again, either one of these capabilities would have been a winner but together they were over the top for a SMEX mission.

So with all of these ``first ever'' capabilities, why did it take so long to get HESSI approved for a SMEX? I have outlined the long and tortuous process that we went through in the last section but when the 1997 SMEX proposal was finally accepted, George Withbroe, the Headquarters selecting official at the time, said it was a ``slam dunk.'' That is a pretty strong statement for a Headquarters guy!

The third and critical reason for RHESSI's success is the team that Bob put together to design, fabricate, and fly the instrument. All the original CoIs on the proposal deserve a mention (in the order listed in the proposal) since each played a crucial role in planning the mission and bringing it to fruition -
Carol Jo Crannell, Gordon Holman, Reuven Ramaty, Tycho von Rosenvinge, Richard Canfield, Gordon Emslie, Hugh Hudson, Gordon Hurford, Norman Madden, Frank van Beek, Arnold Benz, Pat Bornman, John Brown, Shinzo Enome, Takeo Kosugi, Nicole Vilmer, and Alex Zehneder.


And of course there were the many indomitable engineers - Dave Curtis, Dave Pankow, Peter Harvey and Paul Turin at Berkeley, Dave Clark, Rob Boyle, and Stuart Banks at Goddard, Alex Zehnder at the Paul Scherrer Institut (PSI) in Switzerland, and Mike Appleby then with Tecomet and responsible for fabricating the finest grid collimators.

A fourth, and ultimately an equally important, factor responsible for\linebreak RHESSI's success is the data analysis software prepared by a remarkable team of computer experts. It was recognized from the outset that the RHESSI data would be complex. Remarkably, the first Senior Review panel in 2003 recommended that a \$100 K to \$200 K enhancement be made above the in-guide budget ``to support training in data usage and data dissemination.'' 

The software team, led initially by Gordon Hurford but since $\sim$2010 by Richard Schwartz, has done a fantastic job in providing a comprehensive set of software to efficiently and accurately generate light curves, spectra, images, and spatially resolved spectra from the raw data transmitted from the spacecraft. As I understand it, the basic object-oriented IDL scheme was set up originally by one of Arnold Benz's graduate students from Z\"{u}rich, Andre Csillaghy. This required a whole new learning curve for people, even if they were familiar with the standard use of IDL, but it has allowed for the efficient management of the different codes and the execution of them with fairly simple instructions entered at the IDL command line.

For the less computer-ambitious scientists such as myself and many other people both inside and outside the PI team, Kim Tolbert did yeoman service in providing a wonderful graphical user interface (GUI) that we could all use without knowing much about IDL objects or the often arcane IDL commands.  With her help, I wrote two documents called ``RHESSI Imaging - First Steps\footnote{https://hesperia.gsfc.nasa.gov/ssw/hessi/doc/guides/imaging/first\_steps.htm}'' and ``RHESSI Spectroscopy - First Steps''\footnote{https://hesperia.gsfc.nasa.gov/~bdennis/Folders/spectroscopy/first\_steps.htm} that became the standard recipes for many people to get into RHESSI data analysis. They have since become somewhat obsolete but Kim's comprehensive documentation has all the information anybody will ever need to do the most detailed analysis of RHESSI data possible.\footnote{https://hesperia.gsfc.nasa.gov/rhessi3/software/}

But what was my role?  I have struggled to describe my role in this field of thoroughbreds. How did a ``grammar-school boy'' from a farm in England contribute anything useful in the presence of such geniuses in their different fields?  I didn't really have any assigned responsibility other than that of being the official NASA Mission Scientist and the lead CoI at Goddard. 

The Mission Scientist role was somewhat nebulous but basically the idea as I saw it was to monitor progress and work towards getting the best possible scientific return from the mission. Since the PI had the same objectives and we got along so well together, this part of my job turned out to be easy. What could have been a touchy subject was how to divide up the RHESSI funding between Berkeley and Goddard. Again, that was never a bone of contention between the two of us as I know it has been for other missions. 

My role as the Goddard lead CoI was considerably more difficult than being the Mission Scientist. The first and most difficult task, I think, of the whole project was getting the grids fabricated to the required tight specifications set by Gordon Hurford. After much effort, with many failed attempts, we were finally able to obtain flight grids covering the full range of slit widths and pitches needed to allow for both X-ray and gamma-ray imaging with the baseline angular resolution of 2--180 arcsec for X-rays and 36--180 arcsec for gamma-rays. Mike Appleby, then working for Tecomet near Boston, MA, was able to solve the acute problem of making the grids with the finest slits using an etching and stacking technique. Frank van Beek, who had been responsible for the HXIS instrument on SMM, made the coarser tungsten grids and devised the kinematic mounting scheme that ensured that the front and rear grids stayed in exact alignment through launch. Without the efforts of these two men, RHESSI's Fourier-transform imaging technique would not have been possible.

Ultimately, Goddard was responsible for procuring all of the grids, characterizing them, both optically and with X-rays, and working with Alex Zehnder at PSI to install and align them on the end plates of the telescope structure.  This major task was achieved with the dedicated involvement of many people, notably Gordon Hurford, who spent over a year at Goddard supervising this effort; Dave Clark, the mechanical genius who, with his talented hands, did all the mounting and aligning of the grids; Merrick Berg did much of the Optical characterization of the grids using a system set up originally for Carol Crannell to characterize the grids for the HEIDI balloon instrument; and Larry Orwig did much of the X-ray characterization with a system designed and built by Dave Clark.

Goddard was also responsible for providing the cryocooler. Fortunately, people in the Cryogenic Division had been testing a relatively low-cost cryocooler that would satisfy the need to cool the germanium detectors down to temperatures as low as 70~K.  Rob Boyle and Stuart Banks had been able to purchase a large sample of these SunPower coolers and they did yeoman service in providing the flight cooler that has been working flawlessly these last 16 years. This despite the concerns expressed by the expert from JPL, who said at the final go-no-go evaluation meeting that this was a ``tactical cooler'' designed to work for a short time and that we would be lucky if it survived launch. The availability of this cryocooler at such a low cost relative to other coolers with similar capabilities was critical for HESSI to stay within budget.

All I can say about my role is that I did what I could to help all of the scientists and engineers to work efficiently together to make the ``sausage'' that became one of the most successful SMEXes of all time.  It wasn't always pretty but we got the job done despite many setbacks as outlined in the previous section. Perhaps I can take a little pride in my contributions, mostly as a liaison between the experts in the different aspects of what is a fairly complex instrument. As Frank van Beek told me, he thinks of me as the oil on troubled waters, maintaining calm in the presence of the chaos around me.  As an illustration of this role, I well remember a crisis when, to save time, we were simultaneously characterizing and environmentally testing the 36 flight and spare grids at Goddard. Dave Clark had come up with a sequence for doing the grid vibration tests but when Gordon Hurford found out about it, he came storming into my office the next morning and roared that \emph{``This plan is optimally awful.''} It conflicted with his schedule for characterizing the grids. It was left up to me to negotiate with Dave Clark and come up with a more optimal plan, which I did and we got the job done. 

\section{The Future of High Energy Solar Physics}
	\label{Future}
    
\begin{figure}
	\begin{center}
    \includegraphics*[width=0.7\textwidth, angle = 0, 
   trim = 0 0 0 0]
   		{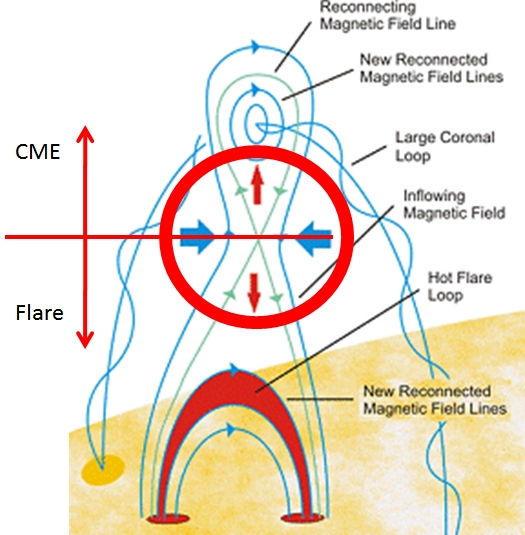}
    \caption{Cartoon of an SEE showing the coronal site of greatest interest inside the red circle where magnetic reconnection is taking place, energy is released, and particles are likely accelerated. }
   \label{Fig-SEECartoon}
   \end{center}
\end{figure}

In the last few years of his life, Bob Lin worked tirelessly to set the direction for the future of high-energy solar physics that would follow on where RHESSI left off. He often told me that he believed that future observations should concentrate on studying all emissions from the region of the coronal HXR sources that RHESSI had detected in many events where the bright footpoints were relatively weaker than usual or were occulted behind the limb. This idea is shown in Figure \ref{Fig-SEECartoon}, where the coronal site of greatest interest is shown in Gordon Holman's cartoon of the initial stages of an SEE. In this simple model, the flare components are assumed to be those that are directed downwards from the magnetic reconnection site and the CME is directed upwards. Bob's idea was that the best information about energy release and particle acceleration could be obtained by new instrumentation optimized to detect all the emissions from this region or from as close to this region as possible. Particularly for electrons, it would no longer be adequate to just detect HXRs from the bright footpoints, the emphasis of all previous instruments since the first detection of footpoint emission by \cite{1981ApJ...246L.155H}. Similarly, all other emissions from the coronal region should also be measured including soft X-rays, gamma-rays, UV, EUV, optical, IR, radio.  The goal would be to determine the spectrum and location of the accelerated particles, both electrons and ions, and the temperature, density, magnetic field, and velocity distributions in this region as functions of time during SEEs.

He laid out his thoughts in his review paper \citep{2011SSRv..159..421L} included at the end of the \textit{Space Science Review} volume on \textit{High Energy Aspects of Solar Flares} that grew out of the annual RHESSI workshops. In the \emph{Future Prospects} section of that paper, he argues that ``Much higher sensitivity HXR imaging spectroscopy is clearly needed to study ... the high coronal sources in typical flares, rather than just the ones for which that source is anomalously bright.''  He also states that ``much larger dynamic range than RHESSI'' is needed ``so that the coronal sources can be observed simultaneously with the very bright footpoint sources.''    Bob believed that these coronal sources were located in the key region where the energy release and particle acceleration takes place.

\subsection{Solar Eruptive Events (SEE) 2020 Mission Concept}
	\label{S-SEE2020}
    
This emphasis on the coronal HXR sources is also evident in the white paper on a SEE mission concept \citep{2013arXiv1311.5243L} that we submitted as one of 288 inputs to the Committee on a Decadal Strategy for Solar and Space Physics (Heliophysics). The full report from this committee \citep{NAP13060} has the same picture of Bob that I have used in Figure \ref{Fig-FrostLinRamaty} with the following caption: ``This report is dedicated to the memory of Robert P. Lin (1942-2012), a pioneering space scientist and a beloved colleague.'' 

The white paper grew out of a community-wide meeting in 2010 that Bob and I organized following the 10th RHESSI annual workshop in Annapolis, MD\footnote{https://rhessi10.wordpress.com/planning-a-new-mission-2/}.  (I made a pitch\footnote{https://hesperia.gsfc.nasa.gov/~ayshih/2020/Annapolis20103.08\_DennisB/} at that meeting for super high-resolution X-Ray Imaging (0.1 arcsec at 6.7 keV) following my work with Gerry Skinner on phase Fresnel lenses and formation flying \citep{2012SoPh..279..573D,2013IJSS&E..1..331S}, and for HXR polarimetry between 10 and 50 keV based on my work with Joe Hill and Kevin Black on Time Projection Chambers \citep{2011ExA....32..101H},  but I guess those ideas were premature as they were not included in the final white paper.  Perhaps they will be relevant for future opportunities one or two solar cycles in the future.) 

\begin{figure}
    \includegraphics*[width=1.0\textwidth, angle = 0, 
   trim = 0 0 0 0]
   		{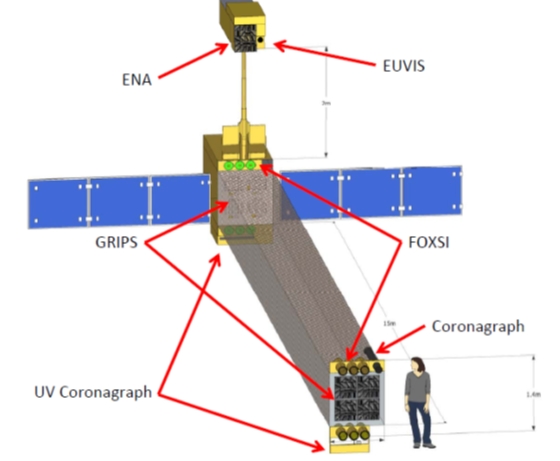}
    \caption{The SEE-2020 mission concept in the white paper. }
   \label{Fig-SEE2020}
\end{figure}

The SEE 2020 white paper laid out a concept for a new mission to advance high-energy solar physics beyond what had been achieved with RHESSI - see Figure \ref{Fig-SEE2020}.
Thanks to Albert Shih, this became known as the \textit{Solar Eruptive Events 2020 Mission Concept} (SEE 2020). (Including ``2020'' in the name was intended to indicate the decade in which it was hoped that it would fly but also that the instruments would have \textit{``20/20 vision.''}) Importantly, it was not just a flare mission but it recognized that most large flares are accompanied by coronal mass ejections with the combination of the two known as a solar eruptive event or SEE.  

The publication of the Solar Flare Myth paper \citep{1993JGR....9818937G} had made it clear back in the '90s that the CME, not the flare was the main geo-effective component of an SEE and the major driver of space weather. However, it seems to me that a CME myth has developed since then in that all you have to worry about is the CME since the associated flare has limited consequences at the Earth or elsewhere in interplanetary space.  This seems to be the case for forecasters of space weather, who have developed detailed models of CMEs and of their passage from the Sun out into space to predict their arrival times at the Earth. Their only interest in the associated flare seems to be if it can tell them anything about the direction of the CME magnetic field that controls the reconnection with the Earth's field. 

A recent paper by \cite{2018arXiv180104608v1G} discusses ``observations and physical mechanisms behind eruptive activity'' and points out the need to ``draw these approaches together.'' Unfortunately, they chose to use the ``general term CME to encapsulate all eruptive activity types'' including ``eruptive flares, filament eruptions, coronal mass ejections, and failed eruptions.'' I think it is better to use a different term - I prefer SEE - to avoid any confusion with what is usually meant by ``CME'' and to differentiate the coronal mass ejection from the flare aspects of these complex events. In any case, it is clear that in order to fully understand these events, one must study both the flare and the CME together as one integrated system of energy release in the solar atmosphere.  That is what the SEE 2020 mission was designed to do. You can't understand the whole complex event without understanding each of the energetically important components both separately and their interdependencies.

The SEE 2020 instruments certainly reflected Bob's bias towards flare observations with particular emphasis on understanding particle acceleration, both of electrons and ions. He still had as his main motivation for studying X-rays and gamma-rays the conclusion he reached in the 1970s that the 10 to 100 keV electrons producing the bremsstrahlung hard X-rays carried the bulk of the total flare energy \citep{1971SoPh...17..412L,1976SoPh...50..153L}. He often quoted this result in his talks, and the SEE 2020 white paper states explicitly that ``the accelerated particles often contain 10-50\% of the total energy released in the flare.''  Also, the Heliophysics Decadal report includes the statement, obviously written by Bob, that ``RHESSI hard X-ray (HXR) imaging-spectroscopy measurements have shown that accelerated electrons often contain about 50 percent of the solar-fare[sic] energy release...'' 

The more recent global energy budgets published by the team led by Gordon Emslie essentially support this conclusion for 38 large solar eruptive events although the order of magnitude uncertainties on most energy estimates of the different components make such a definitively small range problematic \citep{2012ApJ...759...71E}. \cite{2017ApJ...836...17A} have expanded the number of events analyzed to 399 recorded in 3.5 years after the launch of SDO and claim to have achieved closure in terms of balancing the available free energy in the pre-flare magnetic field with the total energy of the different flare and CME components. However, they estimated the total energy in flare-accelerated electrons by invoking a warm-target model that effectively extends the lower energy cutoff down to an average of 6 keV.  In some cases, this leads to an improbably large energy that is significantly above what has ever been detected as an increase in the total radiated energy from the Sun.  Thus, there is still considerable uncertainty in the total energy carried by the flare-accelerated particles but the fact remains that it is a very large fraction of the energy released and that we still don't understand how it happens.

A similar energetics argument can be made for the importance of gamma-ray observations to study the flare-accelerated ions \citep{2011SSRv..159..167V}.  
\cite{2000IAU..195..123R} had shown that for the 19 gamma-ray line flares recorded with the \textit{Gamma-Ray Spectrometer} on SMM, the energy in flare-accelerated ions above 1 MeV was of the same order as the energy in electrons above 20 keV.  We reached the same conclusion from our analysis of the gamma-ray events recorded by RHESSI \citep{2012ApJ...759...71E}. In addition to this energetics argument, the surprising RHESSI result that the gamma-ray sources are sometimes not from the same locations as the X-ray footpoints \citep{2003ApJ...595L..77H,2006SPD....37.2804H,2006ApJ...644L..93H} further reinforces the importance of future gamma-ray imaging spectroscopy. There are clearly many unanswered questions concerning ion acceleration that can best be addressed by studying the gamma-ray emission.

\subsection{FOXSI and GRIPS}
	\label{S-FOXSI_GRIPS}
    
The ideas for advanced new instruments had been percolating for several years, mainly at Berkeley, and they led to two new instrument designs that would have order of magnitude improvements in both X-rays and gamma-rays over RHESSI's capabilities.  These became known as the \textit{Focusing Optics X-ray Solar Imager} (FOXSI)\footnote{http://foxsi.ssl.berkeley.edu/} and the \textit{Gamma-Ray Imager/Polarimeter for Solar flares} (GRIPS)\footnote{http://grips.ssl.berkeley.edu/}.  Both instruments are being actively pursued by many of Bob Lin's former students - Steven Christe and Albert Shih, both now at Goddard, Milo Buitrago Casas at Berkeley, and Lindsay Glesener now at the University of Minnesota - and also by Arnold Benz's former students, S\"{a}m Krucker and Pascal Saint-Hilaire. It is a pleasure to see them all working with such youthful enthusiasm on these exciting new prospects. I help them where I can but the whole proposal preparation process is now so formalized, especially for the SMEX and MIDEX opportunities, that I feel totally out of my depth.

FOXSI was originally designed as a sounding rocket instrument, and three flights have already taken place with a fourth in preparation. It is now in a Phase-A study for a SMEX opportunity.  Both versions use focusing grazing-incidence X-ray optics to directly produce images on the pixelated detectors as opposed to measuring the spatial Fourier components of the source as done by RHESSI. The specialized optics capable of achieving angular resolutions of better than 10 arc-seconds well into the hard X-ray energy range are provided by Brian Ramsey's group at Marshall Space Flight Center. This newly developed direct X-ray imaging technique provides vastly improved sensitivity and dynamic range than was possible with RHESSI at energies from 4 to 15 keV for the rocket version and up to as high as 50 - 70 keV for the SMEX version.

GRIPS is a much more massive instrument and was envisioned originally to be flown on a high-altitude balloon \citep{2012SPIE.8443E..4HS}.  It uses 16 3D position-sensitive germanium detectors to individually record each energy deposition with a spatial resolution of $<0.1~mm^3$ and an energy resolution of a few keV. As with RHESSI, imaging is achieved by modulating the incident flux to determine the spatial Fourier components of the source but in this case, because of the high spatial resolution of the germanium detectors, only a single tungsten multi-pitch rotating modulation collimator is needed, placed 8 meters above the spectrometer. 

\subsection{Heliophysics Decadal Survey}
	\label{S-Decadal_Survey}
The proposed SEE 2020 mission concept was well received by the solar physics subcommittee of the full Heliophysics Decadal Survey committee, officially called the Panel on Solar and Heliospheric Physics and chaired by Dick Mewaldt with Bob Lin as a member. But it quickly became apparent that they could not recommend SEE 2020 with the proposed full complement of instruments simply because it would cost way more than they were eventually told would be available. This was not surprising as the original mission concept was the stereotypical ``Christmas tree'' that you always start out with when you put together everything you would like to include. The total cost was estimated to be well over a billion dollars. Thus, when the detailed negotiations began in the subcommittee to come up with a viable mission that could be recommended to the full committee as part of the decadal plan, Bob decided that only a small subset of the instruments included in the SEE~2020 white paper could be included to bring the costs down to not much more than 500 million dollars. The Aerospace Corporation in El Segundo, CA, was doing the official cost estimates for the different proposed missions, and their methods depended to a large extent on instrument weight. Since GRIPS was by far the heaviest of the proposed instruments, it was the first to go, with Bob arguing that it could be flown on long-duration balloon flights in the Antarctic. In fact, this has already happened in January 2017, thanks to the heroic efforts of the GRIPS team led by Pascal Saint-Hilaire.

After all the studies of the 12 mission concepts proposed for the Decadal Survey were completed, the final version of SEE 2020, now called just \textit{Solar Eruptive Events} (SEE), was the third of six concepts (the only solar mission) that were endorsed by the committee and moved forward into the first stage of the CATE process. CATE stands for \textit{Cost And Technical Evaluation} and is the process used by all the latest decadal surveys to provide an independent cost estimate for proposed missions. SEE was assessed to be one of the more complex missions studied with ``medium'' technical risk and the highest cost (with ``threats'') of \$655 million. The bottom line was that SEE was not recommended for implementation by the Decadal Survey committee. 


The decision of the Heliophysics Decadal Survey Committee was bitterly disappointing after so many people had put so much effort into preparing the SEE-2020 concept. Bob  had pushed it as hard as he could as a member of the Solar panel. Nevertheless, the endorsement as the only solar mission to be selected for the full CATE evaluation was to prove valuable in promoting the individual instruments of the original SEE-2020 concept for opportunities that arose for balloon flights for GRIPS and rocket flights, and now a SMEX possibility for FOXSI. Again, none of this would have happened without Bob's dedicated and tireless efforts. It's amazing to me to note that Bob also submitted two other white papers to this committee - ``Expansion of the Heliophysics Explorer Program,'' where he points out that ``Heliophysics Explorer missions have the best success record in all respects of any space missions,'' and ``\textit{The Multi-Spacecraft Inner Heliosphere Explorer} (HELIX)...a five-spacecraft mission to study large-scale solar transients in the inner heliosphere and their acceleration of particles to high energies.'' I didn't know about these until I read the final committee report.

\subsection{The Standard SEE Model}
	\label{S_Standard_Model}
    
Since the Decadal Survey, the case for future UV and X-ray spectroscopic solar observations has been made in detail by \cite{2016JGRA..12111667H}. He argues that ``As our knowledge about solar activity increases, so does our need for better and more extensive spectroscopic observations.'' ``High-resolution UV and X-ray spectroscopy are important to understanding the origin and evolution of magnetic energy release in the solar atmosphere, as well as the subsequent evolution of
heated plasma and accelerated particles,'' Such spectroscopy provides information on temperatures, densities, line-of-sight velocities, and magnetic field strengths, and also information about the accelerated particles, their propagation and interaction with the thermal plasma. He outlines what is already known about SEEs and describes what he calls the \emph{Standard Model}, both the 2D version and its extension along the SEE arcade to 3D. He points out that simultaneous imaging spectroscopy of the radiation from both thermal plasma and suprathermal electrons is crucial to obtaining an understanding of both plasma heating and
particle acceleration and their evolution.  These observations must be made on the sub-second time scales and arc-second spatial scales of the relevant physical processes that are involved. Much higher dynamic range capability is required in individual X-ray images than the $\sim$10:1~to~50:1 possible with RHESSI for obtaining reliable spectra of thermal and
nonthermal sources at different spatial locations in the field of view. Observations must be made of the emission from plasma with temperatures ranging from $<$10~kK to $>$10~MK for comparison with numerical models that are now making quantitative predictions of the location and dynamic evolution of flare plasma over this temperature range.  

It is exciting to see new instrument designs and combinations of instruments coming together to meet the requirements laid out by \cite{2016JGRA..12111667H}. Hopefully, such a combination of instruments will be selected for flight in time for the start of the next expected increase in solar activity in $\sim2022$. Perhaps I will be able to participate in some way for my fifth solar maximum that is predicted to peak in 2025. I'll only be 85!



\section{Conclusions}
    \label{S-Conclusions}

So what is the point of writing, or reading, this memoir? Is my story in any way relevant in today's world?  Things have changed so much since I came to the US in 1964 but perhaps there is some value in thinking about these changes, particularly as they concern immigration and how to sustain the wonderful international exploration of science that I have had the privilege of pursuing all of my adult life. I certainly benefited from being born on a working farm in England, for while my parents were certainly not part of the gentry, they were landowners and one step up the social ladder from farm workers.  In other ways, I feel that I was discriminated against in England because of the class system there but once I came to America, I benefited from a sort of reverse discrimination because it seems that most Americans are Anglophiles. The open and international nature of scientific research certainly appeals to me, and NASA has provided me with every opportunity to pursue this work.

The details of my zig-zag career through the different institutionalized and cultural biases that exist in both England and the US are not particularly important but here is a summary as I see it:
\begin{itemize}
	\item Born and raised on a working farm in England, 
	\item Passing the 11-plus exam in England to get into a grammar school.
    \item First in my family to go to a university
    \item Not being able to even apply to Oxford or Cambridge University because I did not take a required latin exam.
    \item Getting a PhD at a ``\emph{red-brick}'' English university.
    \item Going down the ``\emph{brain drain}'' to America.
    \item Turned down by the ``\emph{Hoff Circus}'' for a job back in England.
    \item Accepted for similar job with NASA despite not being a US citizen.
    \item Working with a series of great mentors and colleagues at Goddard and around the world on the Orbiting Solar Observatories, the Solar Maximum Mission, and RHESSI.
\end{itemize}

My main takeaway from this is the importance of free and open communications and exchanges with all people around the world - free from government regulations, from stereotyping, and from destructive conflicts of all kinds. Perhaps the main relevance of my story is in regards to the immigrations issue that is currently of such great concern in the US and around the world. I certainly benefited from the essentially free immigration that was possible at the time from England. I like to think that my life in the US has benefited this country. I've paid my taxes every year, brought three wonderful and successful children into the world, who are now all married and paying their taxes too.  I also think I have contributed some to the success of the US space program at a cost to the US tax payer of some \$10 million ($\sim$\$200~K~per~year for 50~years. Quite a bargain, I'd say! But is it right or sustainable for the US to continue to attract people like me with PhDs to leave their home countries and work for all of their productive years in the US? It seems great for the latest ``\emph{America First}'' policy but is it good for the world as a whole or even for the long-term advancement of science?

In my response to a recent NASA request for comments on a proposal for open access to all computer programs, I wrote the following: ``\emph{Given NASA's goal of maximizing science return, the case for open source, as with open data, is self evident.}''  I feel the same way about open borders between nations.  I recognize that this is not politically tenable at present or even desirable given that, by some estimates, 700 million people want to move to the US.  So perhaps the solution is not to actively attract the top people from other countries but rather to encourage and support efforts to make more opportunities available in these other countries.  In my case, wouldn't it have made more sense from England's point of view to let me take the job at Harwell that I had already been offered. I suspect it was the English class system that led the Hoff Board to reject my application because I was a ``\emph{grammar school boy}'' and not one of the privileged few who had gone to Oxford or Cambridge.  Well it was their loss.

I feel no guilt for not returning to England for nationalistic or patriotic reasons or to repay the cost of my free education and upbringing there. I know many colleagues who have returned to their home countries after spending time in the US, some for altruistic reasons but mostly I think because they liked it better there. I have little pride in being \emph{English}, and even after living in America for most of my life, I don't really feel \emph{``American''}. I have always been attracted to the idea that I am truly \emph{``a citizen of the world.''} I still think that is the best way to be, accepting all peoples irrespective of where they were born or raised.

\begin{figure}
    \includegraphics*[width=1.0\textwidth, angle = 0, 
   trim = 0 0 0 0]
   		{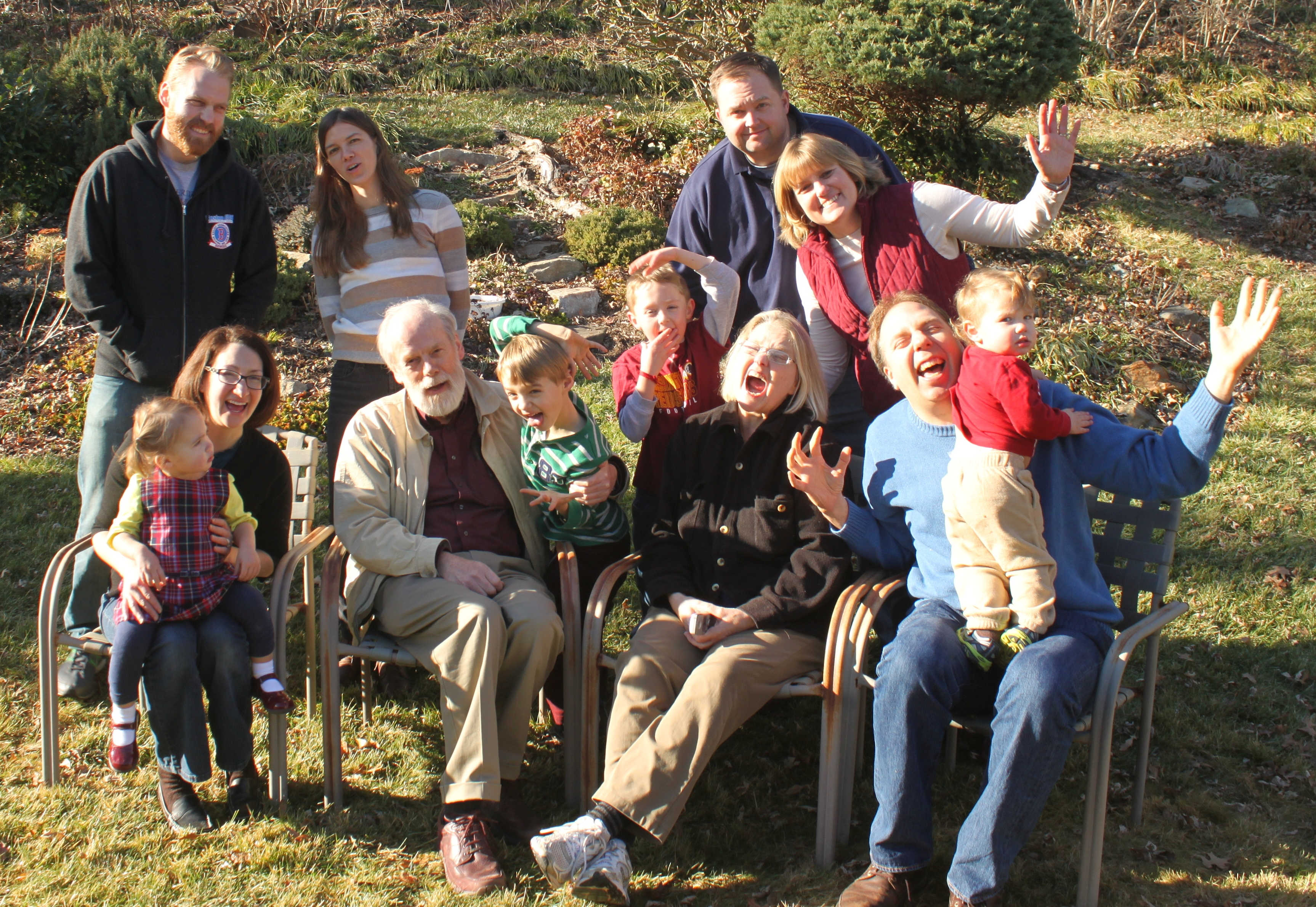}
    \caption{My US family.}
   \label{Fig-Family}
\end{figure}

It has been a privilege to write this memoir. It gives me the opportunity to publicly acknowledge the debt I owe to my many mentors and colleagues who have given of their time that I might better learn the field of solar flare physics and contribute to it in the best way I could.  In particular, I thank Ken Frost for hiring me at Goddard and supporting me for many years until he retired from government service.  Also, I thank Bob Lin for his energy and friendship. Without him, RHESSI would not have happened and I wouldn't have had the wonderful run of interesting and exciting observations for going on 16 years now.  Finally, as I ended my Maniac talk, I must acknowledge my US family in Figure \ref{Fig-Family}, for truly, without them, I would be nothing.

%

%

%


%
\begin{acks}

I acknowledge my debts to so many people that I have interacted with throughout my career. Starting of course, with my parents, who gave me such a wonderful start on a farm in England and supported me all the way until their too early deaths.  I acknowledge all the teachers and school mates at Gipsey Bridge Elementary School, Horncastle Grammar School, and Leeds University. I am indebted to the wonderfully friendly and mostly Anglophile people that have made my life so pleasurable and satisfying in America, first at the University of Rochester and, for over 50 years, at NASA's Goddard Space Flight Center. I also recognize the UK and US taxpayers, for without their financial support I would not have been so well educated and could not have pursued my research work. I hope they got their monies worth. I am so grateful for having the opportunity to participate in the great international effort in pursuit of scientific truth.  I thank unreservedly all fellow scientists and other interested participants who I have had the pleasure of working with on this wonderful journey.  Will it never end? 
Finally, I thank the anonymous referee and the journal editor for actually reading the whole thing and pointing out many errors and improvements that I have attempted to accommodate.

\end{acks}

%
%
%
%
\bibliographystyle{spr-mp-sola}

\bibliography{DennisMemoir22Nov2017}
%
%
%
%

\end{article}
\end{document}